\documentclass[final]{siamart220329} 
\usepackage{amsfonts} 
\usepackage{physics} 
\usepackage{nameref}
\usepackage{mathtools}
\usepackage{tikz} 
\usepackage{tikz-3dplot}
\usetikzlibrary{decorations.markings} 
\usepackage{setspace}
\usepackage{textgreek}

\makeatletter 
\renewcommand\paragraph{
	\@startsection{paragraph}{4}{\z@}
	{-2.5ex\@plus -1ex \@minus -.25ex}
	{1.25ex \@plus .25ex}
	{\normalfont\normalsize\bfseries}
	}
\makeatother 
\crefname{paragraph}{subsection}{subsections}
\Crefname{paragraph}{Subsection}{Subsections}

\newcommand{\R}{\mathbb{R}}
\newcommand{\ds}{\displaystyle}
\newcommand{\pard}[2]{\frac{\partial#1}{\partial#2}}
\newcommand{\pardd}[2]{\frac{\partial^2#1}{\partial#2^2}}

\headers{A Multi-Scale Simulation of Retinal Physiology}{B. Abuelnasr and A. R. Stinchcombe}

\title{A Multi-Scale Simulation of Retinal Physiology\thanks{
\funding{We acknowledge the support of the Natural Sciences and Engineering Research Council of Canada (NSERC): RGPIN-2019-06946 and the Ontario Graduate Scholarship}}}

\author{Belal Abuelnasr\thanks{Department of Mathematics, University of Toronto, Canada
  (\email{belal.abuelnasr@mail.utoronto.ca}, \email{stinch@math.toronto.edu}).}
\and Adam R. Stinchcombe\footnotemark[2]}

\begin{document}

\maketitle

\begin{abstract}
We present a detailed physiological model of the retina that includes the biochemistry and electrophysiology of phototransduction, neuronal electrical coupling, and the spherical geometry of the eye. 
The model is a parabolic-elliptic system of partial differential equations based on the mathematical framework of the bi-domain equations, which we have generalized to account for multiple cell-types.
We discretize in space with non-uniform finite differences and step through time with a custom adaptive time-stepper that employs a backward differentiation formula and an inexact Newton method.
A refinement study confirms the accuracy and efficiency of our numerical method. 
Numerical simulations using the model compare favorably with experimental findings, such as desensitization to light stimuli and calcium buffering in photoreceptors.
Other numerical simulations suggest an interplay between photoreceptor gap junctions and inner segment, but not outer segment, calcium concentration.
Applications of this model and simulation include analysis of retinal calcium imaging experiments, the design of electroretinograms, the design of visual prosthetics, and studies of ephaptic coupling within the retina.
\end{abstract}

\begin{keywords}
Adaptive solver, inexact Newton method, Newton-iterative methods, retina model, simulation, biological model 
\end{keywords}

\begin{MSCcodes}
65N06, 35B27, 92-08, 92-10
\end{MSCcodes}

\section{Introduction}

The retina is a unique part of the brain in that it is physically exposed to the outside world and has a relatively simple anatomical structure as compared to other parts of the brain~\cite{Dowling}.
This explains why it has been heavily studied for well over a century now~\cite{Dowling} and, hence, much about retinal physiology and anatomical structure is well understood.
For example, the anatomical structure and density of photoreceptors, which are the site of light detection and where signal transduction and processing begins~\cite{Rodieck}, are well known in numerous species~\cite{Wikler, Hoang, Fu1, Zang}.
Moreover, much has been discovered about the various processes and functions of the retina.
For instance, phototransduction, which is the process in which light is converted into an electrical signal, is understood in great detail.
There is a widely accepted description of phototransduction including the various molecules and proteins crucial to this process, such as opsins that isomerize after absorbing a photon 
and guanosine 3',5'-cyclic monophosphate (cGMP) that keeps some cGMP-gated cation channels open maintaining a dark current~\cite{Fu2}.
As a result of all this research, many biological models of the various retinal neurons have been developed~\cite{Kamiyama, Usui1, Usui2, Aoyama, Invergo}.

Being a physically accessible part of the brain, there have been numerous experimental studies on the electrical activity in and around the retina as a result of light stimuli.
In particular, visual prosthetics~\cite{Fernandez, ECohen, Weiland} and medical diagnosis~\cite{Creel, Dowling} are two prominent applications of such studies.
Electroretinograms (ERGs) are a clinical diagnosis tool in which patients are shown various flashes of light stimuli and the resulting change in electrical potential on the eye surface is recorded.
ERGs can be used to detect various diseases, such as retinitis pigmentosa and retinal vascular diseases~\cite{Creel}.
A biologically-realistic, full-retina model would aid such studies and their applications.

Considering that the retina is composed of hundreds of millions of neurons, we will use a continuum model in which the electrical activity is described in spatial aggregate rather than in a cell-by-cell basis.
The bi-domain model, first introduced by Tung, is such a continuum model, which uses homogenization to account for the multiple scales present in similar tissue-level models~\cite{Tung}.
This model includes two separate domains (extracellular and intracellular spaces) which occupy the same physical space.
Current is allowed to flow between the two domains (see \cref{bidomain derivation}).
The currents and the dynamic variables in the bi-domain equations are spatial averages, over a large number of cells, of the corresponding currents and dynamic variables for individual cells, derived through homogenization~\cite{Tung, Henriquez}.
Homogenization is a mathematical technique which is used to obtain the macroscopic properties of a system from its microscopic ones~\cite{Henriquez, Cioranescu}.
It can be used to obtain averaged equations from a system of partial differential equations whose (spatial) domain has a periodic microstructure~\cite{Keener}.
See~\cite{Keener, Henriquez} for a derivation of the bi-domain equations using homogenization.

The bi-domain equations have been applied extensively in modelling cardiac tissue.
One such example is in predicting and suggesting mechanisms~\cite{Sepulveda, Roth1, Roth2, Roth3, Roth4, Bennett} for cardiac strength-interval curves~\cite{Kandel}.
Results from studies using the bi-domain equations~\cite{Roth4, Bennett} resembled those obtained from experimental measurements~\cite{Dekker, Mehra}.
In some cases, studies using the bi-domain equations~\cite{Sepulveda} predicted mechanisms that were only later confirmed by experimental studies~\cite{Wikswo1, Knisley1, Neunlist1, Tung1, Neunlist2, Wikswo2, Knisley2}.
We hope to attain similar results in modeling the retina using the bi-domain equations.

The bi-domain equations have also been used to study neural tissues~\cite{Nagarajan, Sadleir1, Sadleir2, Meffin}.
The most notable bi-domain model of the retina was proposed by Dokos et al.~\cite{Dokos}, which is concerned with epiretinal stimulation via stimulus electrodes.
This model has since been further developed, in various directions, for various purposes, such as handling different electrode stimulation techniques, using finite element implementation, and incorporating more details of the retina~\cite{Joarder1, Yin1, Yin2, Joarder2, Shalbaf1, AlAbed, Alqahtani1}.  
Numerous studies have been conducted based on the various versions of this model and have largely been concerned with visual prosthetics and/or electrical stimulation of the retina~\cite{Abramian, Shalbaf2, Shalbaf3, Alqahtani2, Alqahtani3}.
Given the scope of their work, the model proposed by Dokos et al. does not accurately describe the entire geometry of the retina and makes no attempt to model the entirety of the vitreous chamber.
Also, the stimuli to the retina in most versions of this model was provided solely by stimulus electrodes.
Only one version of this model was developed with light stimuli, but without detailed biological descriptions of some of the neurons, including photoreceptors, nor did they account for the geometry of the eye~\cite{Yin2}.
In particular, they studied the retinal response to small and large light-spot stimuli.
Their findings were consistent with experimental studies, especially as it pertains to the relation between the size of the stimuli and surround antagonism~\cite{Yin2}.

An important reason for using the bi-domain equations in modeling the retina (and cardiac tissue) is that it provides an accurate description of the micro-scale structures in a macro-scale model.
To illustrate, tissues are made out of cells on the micro-scale, and the intracellular and extracellular spaces are physically separated by cell membranes. 
Using the bi-domain equations, we retain this micro-scale description in the macro-scale model.
However, when modeling tissues with multiple, densely packed cell-types, such as the four known photoreceptors of the retina~\cite{Dowling, Rodieck, Kolb2}, the presence of the different intracellular spaces is not addressed in the macro-scale model.
For this reason, we generalized our model to handle this multi-domain scenario (see \cref{multidomain derivation} for the derivation of the multi-domain equations).
This generalization also allows us to apply different light stimuli for the different photoreceptors.
Thus, we are able to account for the differences in the sensitivities of the various photoreceptors to light of certain wavelengths~\cite{Dowling, Rodieck, Kolb2}.
A dissimilar multi-domain model of the retina, based on Dokos et al., has been proposed~\cite{Alqahtani1}.
The multiple domains in that model represent the different compartments of the retinal ganglion cells, such as the dendrites, soma and axon, rather than different cell-types~\cite{Alqahtani1, Alqahtani2, Alqahtani3}.

We present a detailed model of the retina, which takes into account retinal physiology and the spherical geometry of the eye.
Light incident on the retina provides the stimuli, through a model of the phototransduction pathway~\cite{Kamiyama}.
The retina model is a system of PDEs which we solve using a finite difference scheme (see \cref{bidomain derivation,spatial discretization}). 
We overcome many challenges to successfully model the retina in this way including the 3D nature of this problem, the spherical geometry of the eye, numerical stiffness of the retinal dynamics, and the multiple scales involved in this model.
We implicitly step through time using a backwards differentiation formula and Newton's method (see \cref{time-stepping}).
We present an adaptive time-stepper (\cref{adaptive time-stepping}) and an inexact Newton method (\cref{Jacobian update}) to mitigate the computational cost and time arising from such complications.

We discuss the details of our simulation in \cref{methods}.
We begin by describing the spatial aspect of the model (\cref{sec: geometry}).
We then discuss the mathematical basis of the model, including a derivation of the bi-domain and multi-domain equations and the additional assumptions we make (\cref{model derivation}).
This is followed by a detailed discussion of the numerical methods we used to solve our system of PDEs (\cref{numerical methods}).
Details from the spatial discretization (\cref{spatial discretization}) and the implicit time-stepping scheme used (\cref{time-stepping}), to the adaptive time-stepper (\cref{adaptive time-stepping}) and the inexact Newton method used (\cref{Jacobian update}) are included in that discussion.
Subsequently, we present and discuss findings (\cref{results and discussion}) obtained using a few numerical simulation of the model (\cref{appendix}).
We also present and discuss a convergence study of our implicit time-stepping scheme (\cref{convergence study}), and analyses of the adaptive time-stepper (\cref{nonadaptive-adaptive comparison,adaptive fine details}) and the inexact Newton method used (\cref{direct-iterative comparison}).

\section{Methods}\label{methods}

\subsection{Geometrical Setup}\label{sec: geometry}

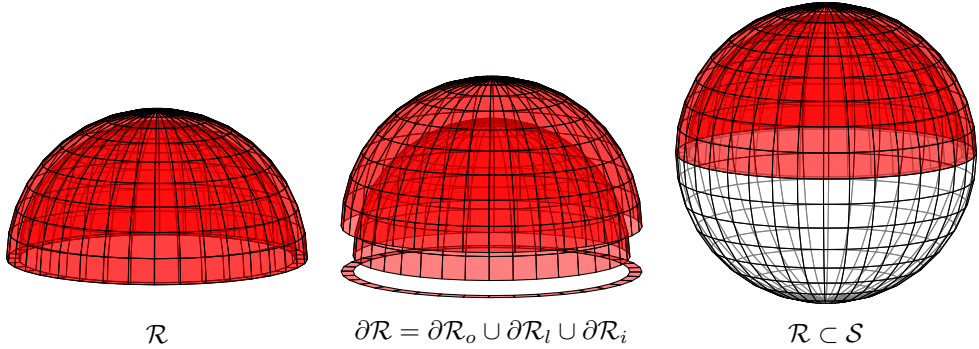
\begin{figure}
 \vspace{0.5 cm}
 \tdplotsetmaincoords{80}{1}
 \begin{tikzpicture}[scale=2,line join=bevel, tdplot_main_coords, fill opacity=0.5]
      \tdplotsetpolarplotrange{0}{90}{0}{360}
       \tdplotsphericalsurfaceplot{36}{36}
      {0.92}{black}{red}
      {}{}{}
      \foreach \theta in {0,10,...,350}
      	\draw[fill = red] 
  		({cos(\theta)*0.92},{sin(\theta)*0.92},{0})
		-- ({cos(\theta+10)*0.92},{sin(\theta+10)*0.92},{0})
		-- ({cos(\theta+10)},{sin(\theta+10)},{0})
		-- ({cos(\theta)},{sin(\theta)},{0})
		-- cycle;  
      \tdplotsphericalsurfaceplot{36}{36}
      {1}{black}{red}
      {}{}{}	
      \node[tdplot_screen_coords,fill opacity=1] at (0,-0.5) {$\mathcal{R}$};
\end{tikzpicture}
\tdplotsetmaincoords{80}{1}
\begin{tikzpicture}[scale=2,line join=bevel, tdplot_main_coords, fill opacity=0.5]
	\tdplotsetpolarplotrange{0}{90}{0}{360}
	\tdplotsphericalsurfaceplot{36}{36} 
            {0.92}{black}{red}
            {}{}{}
	\begin{scope}[yshift = -0.1cm]
            \foreach \theta in {0,10,...,350}
      		\draw[fill = red] 
                    ({cos(\theta)*0.92},{sin(\theta)*0.92},{0})
                        -- ({cos(\theta+10)*0.92},{sin(\theta+10)*0.92},{0})
                        -- ({cos(\theta+10)},{sin(\theta+10)},{0})
                        -- ({cos(\theta)},{sin(\theta)},{0})
                        -- cycle; 
	\end{scope}      
	\begin{scope}[yshift = 0.2cm]            
            \tdplotsetpolarplotrange{0}{90}{0}{360}		 
            \tdplotsphericalsurfaceplot{36}{36}
                {1}{black}{red}
                {}{}{}
	\end{scope}            	
	\node[tdplot_screen_coords,fill opacity=1] at (0,-0.5) {$\partial\mathcal{R} = \partial\mathcal{R}_o \cup \partial\mathcal{R}_l \cup \partial\mathcal{R}_i$};
\end{tikzpicture}
\tdplotsetmaincoords{80}{1}
 \begin{tikzpicture}[scale=2,line join=bevel, tdplot_main_coords, fill opacity=0.5]
	\begin{scope}
          \tdplotsetpolarplotrange{0}{90}{0}{360}
           \tdplotsphericalsurfaceplot{36}{36}
              {0.92}{black}{red}
              {}{}{}
	\end{scope}    
      \foreach \theta in {0,10,...,350}
      	\draw[fill = red] 
  		({cos(\theta)*0.92},{sin(\theta)*0.92},{0})
		-- ({cos(\theta+10)*0.92},{sin(\theta+10)*0.92},{0})
		-- ({cos(\theta+10)},{sin(\theta+10)},{0})
		-- ({cos(\theta)},{sin(\theta)},{0})
		-- cycle;
      \tdplotsetpolarplotrange{90}{180}{0}{360}		
      \tdplotsphericalsurfaceplot{36}{36}
      {1}{black}{white}
      {}{}{}
      \tdplotsetpolarplotrange{0}{90}{0}{360}      
      \tdplotsphericalsurfaceplot{36}{36}
      {1}{black}{red}
      {}{}{}	
      \node[tdplot_screen_coords,fill opacity=1] at (0,-1.2) {$\mathcal{R}\subset\mathcal{S}$};
\end{tikzpicture}
\caption{Left: The retina, $\mathcal{R}$ is shown. Middle: The outer boundary, $\partial\mathcal{R}_{\mathrm{o}}$, lateral boundary, $\partial\mathcal{R}_{\ell}$, and inner boundary, $\partial\mathcal{R}_{\mathrm{i}}$ of the retina are shown. Right: The retina is centered across the north pole of the eye, $\mathcal{S}$. In this set up the cornea would be located around the south pole of the eye.}
\label{geometry}
\end{figure}

We assume the eye $\mathcal{S} = \overline{B(0,r_{\text{eye}})} \subset \R^3$ to be a closed ball centered at the origin with radius $r_\text{eye} = 12.25 \text{ mm}$~\cite{Kolb3}.
We orient the eye so that the retina, $\mathcal{R} = \{(r,\theta,\varphi) \in \mathcal{S} : r_\text{eye}-r_\text{retina}\le r \le r_\text{eye}, \ \varphi_\text{retina}\le \varphi \le \frac{\pi}{2} \}$, is situated on the north pole of $\mathcal{S}$.
We choose $\varphi_\text{retina}=\frac{\pi}{2}$ rad to obtain an experimentally accepted (outer) retinal radius value of $19.2$ mm~\cite{Kolb4}.
We segment the retinal boundary, $\partial\mathcal{R}$, into an outer boundary, $\partial\mathcal{R}_{\mathrm{o}}$, lateral boundary, $\partial\mathcal{R}_{\ell}$, and an inner boundary, $\partial\mathcal{R}_{\mathrm{i}}$ as shown in \cref{geometry}. 
Hence, $\partial\mathcal{R} = \partial\mathcal{R}_{\mathrm{o}} \cup \partial\mathcal{R}_{\ell} \cup \partial\mathcal{R}_{\mathrm{i}}$, with $\partial\mathcal{R}_{\mathrm{o}} = \{(r,\theta,\varphi) \in \mathcal{S} : r = r_\text{eye}, \ \varphi_\text{retina}\le \varphi \le \frac{\pi}{2}  \}, \ \partial\mathcal{R}_{\ell} = \{(r,\theta,\varphi) \in \mathcal{S} : r_\text{eye}-r_\text{retina}\le r \le 1, \ \varphi =\varphi_\text{retina} \}, \ \partial\mathcal{R}_{\mathrm{i}} = \{(r,\theta,\varphi) \in \mathcal{S} : r = r_\text{eye}-r_\text{retina}, \ \varphi_\text{retina}\le \varphi \le \frac{\pi}{2}  \}$.

\subsection{Model Derivation}\label{model derivation}
As shown in \cref{geometry}, we divide the eye into the sensory part, the retina $\mathcal{R}$, and the rest of the eye, $\mathcal{S} \setminus \mathcal{R}$, which will consist of various parts including the vitreous chamber and the lens, assumed to be homogeneous.
We use the bi-domain/multi-domain equations to model the retina, while we model the passive region using the mono-domain equation.
The interface between the two regions is the lateral boundary, $\partial\mathcal{R}_{\ell}$, and the inner boundary $\partial\mathcal{R}_{\mathrm{i}}$.

\subsubsection{Bi-domain Equations Setup}\label{bidomain derivation}

For the sake of completeness, we present a derivation of the bi-domain equations.
Many similar derivations can be found in the literature, for example \cite{Keener}. 
Let $\phi_{\mathrm{i}}, \phi_{\mathrm{e}}$ be the intracellular and the extracellular potentials of the retina, respectively, and $\phi_\mathrm{s}$ be the potential of the rest of the eye. 
Using the microscopic version of Ohm's law we get
\[
J_\mathrm{i} = \mu_{\mathrm{i}}E_\mathrm{i} = -\mu_{\mathrm{i}}\nabla\phi_{\mathrm{i}}, 
\hspace{1cm}
J_\mathrm{e} = \mu_{\mathrm{e}}E_\mathrm{e} = -\mu_{\mathrm{e}}\nabla\phi_{\mathrm{e}},
\hspace{1cm}
J_\mathrm{s} = \mu_{\mathrm{s}}E_\mathrm{s} = -\mu_{\mathrm{s}}\nabla\phi_{\mathrm{s}},
\]
in which $\mu_{\mathrm{i}},\mu_{\mathrm{e}}$ are the conductivity of the intracellular and extracellular domains of the retina, and $\mu_{\mathrm{s}}$ is the conductivity of the interior of the eye; $J_\mathrm{i}, J_\mathrm{e}$ denote the intracellular and extracellular current densities in the retina, and $J_\mathrm{s}$ the current density in the rest of the eye; and $E_\mathrm{i}, E_\mathrm{e}$ denote the intracellular and extracellular electric fields in the retina, and $E_\mathrm{s}$ the electric field in the rest of the eye.
In our setting, the conductivities will be (symmetric positive definite) tensors that are functions of time as discussed in \cref{spatial discretization}.
As is typical in electrostatics, our assumption that charge cannot accumulate at any point on the passive region of the eye takes the form
\[
\nabla\cdot J_\mathrm{s} = 0.
\]
As each point in the retina resides in both the intracellular and extracellular domain, our assumption about charge accumulation takes the form
\[
\nabla\cdot (J_\mathrm{e}+J_\mathrm{i})=0.
\]
In each domain,  transmembrane currents, capacitive currents, and any applied currents appear as sources
\begin{align*}
\nabla\cdot J_\mathrm{i} &= -\nabla\cdot(\mu_{\mathrm{i}}\nabla\phi_{\mathrm{i}}) = -\frac{1}{\chi}(C_{\mathrm{m}} \pard{V_{\mathrm{m}}}{t} + I_{\mathrm{m}}), \\
\nabla\cdot J_\mathrm{e} &= -\nabla\cdot(\mu_{\mathrm{e}}\nabla\phi_{\mathrm{e}}) = \frac{1}{\chi}(C_{\mathrm{m}} \pard{V_{\mathrm{m}}}{t} + I_{\mathrm{m}}),
\end{align*}
in which $\chi$ is the volume-to-surface ratio of the cell membrane, $C_{\mathrm{m}}$ is the cell membrane capacitance, $I_{\mathrm{m}}$ is the transmembrane current density, and $V_{\mathrm{m}} = \phi_{\mathrm{i}} - \phi_{\mathrm{e}}$ is the membrane voltage, the difference between the intracellular and extracellular potential.
The sign difference between the right hand sides is expected since current exiting the intracellular region enters the extracellular region.

We choose the transmembrane currents to be given by a conductance-based, rod-photoreceptor model~\cite{Kamiyama}, the Kamiyama model.
However, our model is adaptable to any choice of transmembrane currents.
In transmembrane current models where the total current is reported rather than the density, one needs to divide by the surface area of the cell membrane, which may be absorbed into $\chi$.
Parameters associated with the transmembrane currents are permitted to depend on space if required.
Since cone and rod photoreceptors have similar ionic currents in their inner segments~\cite{Barnes}, we can also model cones by reducing some time constants in the Kamiyama model.
As photoreceptors are the main retinal cells of our interest, we used the same set of equations for the transmembrane currents throughout the entire retina.

To address boundary conditions of this system, we start by assuming that the eye is surrounded by perfectly insulating material (so that the current cannot leave the eye)
\begin{align*}
n_\mathrm{o} \cdot (\mu_{\mathrm{i}} \nabla \phi_{\mathrm{i}}) &= 0 & \text{(on $\partial\mathcal{R}_\mathrm{o}$)},\\
n_\mathrm{o} \cdot (\mu_{\mathrm{e}} \nabla \phi_{\mathrm{e}}) &= 0 & \text{(on $\partial\mathcal{R}_\mathrm{o}$)},\\
n_\mathrm{s} \cdot (\mu_{\mathrm{s}} \nabla \phi_{\mathrm{s}}) &= 0 & \text{(on $\partial\mathcal{S} \setminus \partial\mathcal{R}_\mathrm{o}$)},
\end{align*}
in which $n_\mathrm{s}$ is the normal vector to $\partial\mathcal{S} = \{(r,\theta,\varphi) \in \mathcal{S} : r=r_\text{eye} \}$ and $n_\mathrm{o}$ is the normal to $\partial\mathcal{R}_\mathrm{o}$.
On the boundary of the retina and the rest of the eye, where the transition of a bi-domain to a mono-domain occurs, we require that the extracellular potential and the current be continuous
\begin{align}
 \phi_{\mathrm{e}} &= \phi_{\mathrm{s}}	& \text{(on $\partial\mathcal{R}_\mathrm{i} \cup \partial\mathcal{R}_\ell$)},\label{eq: potential continuity}\\
n_\mathrm{x} \cdot (\mu_{\mathrm{i}} \nabla \phi_{\mathrm{i}} + \mu_{\mathrm{e}} \nabla \phi_{\mathrm{e}}) &= n_\mathrm{x} \cdot (\mu_{\mathrm{s}} \nabla \phi_{\mathrm{s}}) & \text{(on $\partial\mathcal{R}_\mathrm{x}$ for $\mathrm{x} = \mathrm{i}, \ell$)}.\nonumber
\end{align}
Following Tung~\cite{Tung}, we make the additional assumption that the intracellular domain of the retina is isolated from the rest of the eye, which gives
\begin{align}
n_\mathrm{x} \cdot (\mu_{\mathrm{i}} \nabla \phi_{\mathrm{i}}) &= 0 & \text{(on $\partial\mathcal{R}_\mathrm{x}$ for $\mathrm{x} = \mathrm{i}, \ell$)}, \nonumber\\
n_\mathrm{x} \cdot (\mu_{\mathrm{e}} \nabla \phi_{\mathrm{e}}) &= n_\mathrm{x} \cdot (\mu_{\mathrm{s}} \nabla \phi_{\mathrm{s}}) & \text{(on $\partial\mathcal{R}_\mathrm{x}$ for $\mathrm{x} = \mathrm{i}, \ell$)}\label{eq: curr continuity}.
\end{align}

In summary, our model is a system of partial differential equations for unknowns $\phi_{\mathrm{i}}, \phi_{\mathrm{e}}$ (defined on $\mathcal{R}$), and $\phi_{\mathrm{s}}$ (defined $\overline{\mathcal{S}\setminus \mathcal{R}}$),
\[
\text{retina}\left\{
\begin{array}{ll}
\nabla \cdot (\mu_{\mathrm{i}} \nabla \phi_{\mathrm{i}} + \mu_{\mathrm{e}} \nabla \phi_{\mathrm{e}}) = 0 \hspace{7.9 em}  &\text{(on $\mathcal{R}$),} \\
\nabla \cdot (\mu_{\mathrm{e}} \nabla \phi_{\mathrm{e}}) = -\frac{1}{\chi}\left(C_{\mathrm{m}} \frac{\partial V_{\mathrm{m}}}{\partial t} + I_{\mathrm{m}}(V_\mathrm{m}, \boldsymbol{X})\right) & \text{(on $\mathcal{R}$),}\\
V_{\mathrm{m}} = \phi_{\mathrm{i}} - \phi_{\mathrm{e}}  &\text{(on $\mathcal{R}$),} \\
\frac{\partial \boldsymbol{X}}{\partial t} = \boldsymbol{G}(V_\mathrm{m},\boldsymbol{X}) & \text{(on $\mathcal{R}$),}\\
\end{array}
\right.
\]
\[
\text{retina boundary}\left\{
\begin{array}{ll}
n_\mathrm{x} \cdot (\mu_{\mathrm{i}} \nabla \phi_{\mathrm{i}}) = 0 \hspace{9.5 em} & \text{(on $\partial\mathcal{R}_\mathrm{x}$ for $\mathrm{x} = \mathrm{o}, \mathrm{i}, \ell$),} \\
n_\mathrm{o} \cdot (\mu_{\mathrm{e}} \nabla \phi_{\mathrm{e}}) = 0 & \text{(on $\partial\mathcal{R}_\mathrm{o}$),} \\
 \phi_{\mathrm{e}} = \phi_{\mathrm{s}}	& \text{(on $\partial\mathcal{R}_{\mathrm{i}} \cup \partial\mathcal{R}_\ell$),} \\
n_\mathrm{x} \cdot (\mu_{\mathrm{e}} \nabla \phi_{\mathrm{e}}) = n_\mathrm{x} \cdot (\mu_{\mathrm{s}} \nabla \phi_{\mathrm{s}}) & \text{(on $\partial\mathcal{R}_\mathrm{x}$ for $\mathrm{x} = \mathrm{i}, \ell$),}
\end{array} 
\right.
\]
\[
\text{rest of the eye}\left\{
\begin{array}{ll}
\nabla \cdot (\mu_{\mathrm{s}} \nabla \phi_{\mathrm{s}}) = 0 \hspace{12.4 em} & \text{(on $\mathcal{S} \setminus \mathcal{R}$),} \\
n_\mathrm{s} \cdot (\mu_{\mathrm{s}} \nabla \phi_{\mathrm{s}}) = 0 & \text{(on $\partial\mathcal{S} \setminus \partial\mathcal{R}_\mathrm{o}$),}
\end{array}
\right.
\]
in which $t$ is time, $n_\mathrm{s}$ is the outward normal vector to $\partial\mathcal{S}$, $n_\mathrm{x}$ is the normal to $\partial\mathcal{R}_\mathrm{x}$ (pointing away from the retina) for $\mathrm{x} = \mathrm{o}, \mathrm{i}, \ell$, $\boldsymbol{X}$ contains all of the auxiliary dynamic variables from the transmembrane currents model (including channel gating variables and concentrations of proteins involved in phototransduction), and $\boldsymbol{G}$ describes their evolution in time.

We eliminate $\phi_{\mathrm{i}}$ and re-arrange for a dynamic equation for $V_{\textrm{m}}$,
\begin{align}
&\nabla \cdot (\mu_{\mathrm{s}} \nabla \phi_{\mathrm{s}}) = 0 & \text{(on $\mathcal{S} \setminus \mathcal{R}$),} \label{eq: bidomain diffusion}\\
&\nabla \cdot (\mu_{\mathrm{i}} \nabla (V_{\mathrm{m}} + \phi_{\mathrm{e}}) + \mu_{\mathrm{e}} \nabla \phi_{\mathrm{e}}) = 0  &\text{(on $\mathcal{R}$),} \label{eq: bidomain conservation}\\
&\frac{\partial V_{\mathrm{m}}}{\partial t} = -\frac{1}{C_{\mathrm{m}}}\left(\chi \nabla \cdot (\mu_{\mathrm{e}} \nabla \phi_{\mathrm{e}}) + I_{\mathrm{m}}(V_\mathrm{m},\boldsymbol{X})\right) & \text{(on $\mathcal{R}$)}, \label{eq: bidomain voltage}\\
&\frac{\partial \boldsymbol{X}}{\partial t} = \boldsymbol{G}(V_\mathrm{m},\boldsymbol{X}) & \text{(on $\mathcal{R}$)} \label{eq: bidomain auxiliary}.
\end{align}
\Cref{eq: bidomain diffusion,eq: bidomain conservation} are the elliptic constraints on the evolution of the system in time.
In \cref{spatial discretization}, we see explicitly that the discretization of this PDE system is a system of differential algebraic equations (DAE). The elliptic constraints become the algebraic restriction on the system.

\subsubsection{Uniqueness and Boundary Conditions}\label{uniqueness}
By inspecting the system of partial differential equations, it is clear that its solutions are not unique.
That is, if $V_\mathrm{m},\phi_\mathrm{e},\phi_\mathrm{s}$ are solutions then so are $V_\mathrm{m},\phi_\mathrm{e} + c,\phi_\mathrm{s} + c$ for any constant $c$.
Potentials are not unique so long as the potential difference is unchanged.
Since we will be solving this system in time as well, it is important to note that $c$ is allowed to depend on time.
We choose $c$ to ensure that 
\begin{equation}\label{grounding at infinity}
\int_{\partial\mathcal{S}\setminus \partial\mathcal{R}} \phi_\mathrm{s} + \int_{\partial\mathcal{R}_\mathrm{o}}\phi_\mathrm{e} = 0
\end{equation} 
at all times.
The motivation behind this choice can be thought of as selecting the ground electrical potential to be zero.
Let $\phi_\mathrm{o}$ be the harmonic function on $\R^3\setminus \mathcal{S}$ whose values on $\partial\mathcal{S}$ matches $\phi_\mathrm{s}$ and $\phi_\mathrm{e}$, and let $\phi_\infty = \ds\lim_{r\to\infty} \phi_\mathrm{o}$.
Hence, $\phi_\mathrm{o}$ is an extension of $\phi_\mathrm{e}$ (on $\partial\mathcal{R}_\mathrm{o}$) and $\phi_\mathrm{s}$ (on $\partial\mathcal{S}\setminus \partial\mathcal{R}$), while $\phi_\infty$ can be thought of as the ground potential (the potential at infinity).
Using the mean value property on the Kelvin transform of $\phi_\mathrm{o}$ shows that our assumption is tantamount to $\phi_\infty = \int_\mathcal{\partial S} \phi_\mathrm{o} =0$.

\subsubsection{Multi-domain Equations Setup}\label{multidomain derivation}
We will derive the multi-domain equations with an arbitrary number of distinct intracellular domains, say there are $q$ of them with intracellular potentials $\phi_\mathrm{i}^1, \phi_\mathrm{i}^2, \ldots, \phi_\mathrm{i}^q$. 
The derivation will be very similar to that of the bi-domain equations (\cref{bidomain derivation}), so many of the details will be omitted.
We also omit redefining variables introduced previously.
Considering the multi-domain setup, assuming that charge cannot accumulate at any point gives
\[
\nabla \cdot(J_\mathrm{e}+\sum_{j = 1}^{q}{J_\mathrm{i}^j}) = 0,
\]
in which $J_\mathrm{i}^j$ is the current density of the $j$th intracellular domain.
Different photoreceptors are electrically coupled via channels called gap junctions~\cite{Dowling, Rodieck, Kolb2}.
Hence, gap junctional currents also appear as a source of current
\[
\nabla \cdot J_\mathrm{i}^j = -\nabla \cdot (\mu_\mathrm{i}^j\nabla\phi_\mathrm{i}^j) = -\frac{1}{\chi^j}\left(C_\mathrm{m}^j \frac{\partial V_\mathrm{m}^j}{\partial t} + I_\mathrm{m}^j + \sum_{k=1}^q g_{jk}(\phi_\mathrm{i}^j - \phi_\mathrm{i}^k)\right),
\]
in which $\mu_\mathrm{i}^j$ is the conductivity of the $j$th intracellular domain, $\chi^j$ is the volume-to-surface ratio of the cell membrane of the $j$th cell-type, $C_\mathrm{m}^j$ is the cell membrane capacitance of the $j$th cell-type, $I_\mathrm{m}^j$ is the transmembrane current density of the $j$th cell-type, $V_\mathrm{m}^j = \phi_\mathrm{i}^j - \phi_\mathrm{e}$ is the difference of the $j$th intracellular domain potential and extracellular potential, and $g_{jk}$ is the gap junctional conductance between the $j$th and $k$th intracellular domains (we set $g_{jj}=0$, for $j=1, \ldots, q$).
The boundary conditions are handled similarly to the bi-domain equations.
The overall multi-domain model will also be a system of partial differential equations for unknowns $\phi_\mathrm{i}^j,\phi_\mathrm{e}$, and $\phi_\mathrm{s}$ (for $j=1,\ldots,q$)).
These are
\[
\text{\rotatebox{90}{\kern-1em retina}}\left\{
\begin{array}{ll}
\nabla \cdot(\mu_\mathrm{e}\nabla \phi_\mathrm{e}+\sum_{j = 1}^{q}{\mu_\mathrm{i}^j\nabla \phi_\mathrm{i}^j}) = 0 \hspace{10 em}  &\text{(on $\mathcal{R}$),} \\
\nabla \cdot (\mu_\mathrm{i}^j\nabla\phi_\mathrm{i}^j) = \frac{1}{\chi^j}(C_\mathrm{m}^j \frac{\partial V_\mathrm{m}^j}{\partial t} + I_\mathrm{m}^j + \sum_{k=1}^q g_{jk}(\phi_\mathrm{i}^j - \phi_\mathrm{i}^k)) & \text{(on $\mathcal{R}, \ j=1, \ldots, q$),}\\
V_{\mathrm{m}}^j = \phi_{\mathrm{i}}^j - \phi_{\mathrm{e}} \hspace{8.4 em}  &\text{(on $\mathcal{R}, \ j=1, \ldots, q$),}\\
\frac{\partial \boldsymbol{X}^j}{\partial t} = \boldsymbol{G}^j(V_\mathrm{m}^j,\boldsymbol{X}^j) & \text{(on $\mathcal{R}, \ j=1, \ldots, q$),}\\
\end{array}
\right.
\]
\[
\text{\rotatebox{90}{\kern-3.5em retina boundary}}\left\{
\begin{array}{ll}
n_\mathrm{x} \cdot (\mu_{\mathrm{i}}^j \nabla \phi_{\mathrm{i}}^j) = 0 \hspace{9.5 em} & \text{(on $\partial\mathcal{R}_\mathrm{x}$ for $\mathrm{x} = \mathrm{o}, \mathrm{i}, \ell ; \ j=1, \ldots, q$),}\\
n_\mathrm{o} \cdot (\mu_{\mathrm{e}} \nabla \phi_{\mathrm{e}}) = 0 & \text{(on $\partial\mathcal{R}_\mathrm{o}$),} \\
 \phi_{\mathrm{e}} = \phi_{\mathrm{s}}	& \text{(on $\partial\mathcal{R}_{\mathrm{i}} \cup \partial\mathcal{R}_\ell$),} \\
n_\mathrm{x} \cdot (\mu_{\mathrm{e}} \nabla \phi_{\mathrm{e}}) = n_\mathrm{x} \cdot (\mu_{\mathrm{s}} \nabla \phi_{\mathrm{s}}) & \text{(on $\partial\mathcal{R}_\mathrm{x}$ for $\mathrm{x} = \mathrm{i}, \ell$),}
\end{array} 
\right.
\]
\[
\text{\kern-9.2em rest of the eye}\left\{
\begin{array}{ll}
\nabla \cdot (\mu_{\mathrm{s}} \nabla \phi_{\mathrm{s}}) = 0 \hspace{4.8 em} & \text{(on $\mathcal{S} \setminus \mathcal{R}$),} \\
n_\mathrm{s} \cdot (\mu_{\mathrm{s}} \nabla \phi_{\mathrm{s}}) = 0 & \text{(on $\partial\mathcal{S} \setminus \partial\mathcal{R}_\mathrm{o}$),}
\end{array}
\right.
\]
in which $\boldsymbol{X}^j$ contains all of the auxiliary dynamic variables from the transmembrane currents model for the $j$th cell type and $\boldsymbol{G}^j$ describes their evolution in time.

As before, we end up with the following PDE system:
\begin{align}
\nabla \cdot (\mu_{\mathrm{s}} \nabla \phi_{\mathrm{s}}) = 0 &&\text{(on $\mathcal{S} \setminus \mathcal{R}$),} \label{eq: multidomain diffusion}\\
\nabla \cdot\left(\mu_\mathrm{e}\nabla \phi_\mathrm{e}+\sum_{j = 1}^{q}{\mu_\mathrm{i}^j\nabla (V_\mathrm{m}^j+\phi_\mathrm{e})}\right) = 0  &&\text{(on $\mathcal{R}$),} \label{eq: multidomain conservation}\\
\frac{\partial V_{\mathrm{m}}^j}{\partial t} = \frac{1}{C_{\mathrm{m}}^j}\left(\chi^j \nabla \cdot \left(\mu_\mathrm{i}^j\nabla (V_\mathrm{m}^j+\phi_e)\right)\phantom{\sum_{k=1}^q \sum_{k=1}^q}\right. &&\label{eq: multidomain voltage} \\
 \left.- I_{\mathrm{m}}^j(V_\mathrm{m},\boldsymbol{X}^j)-\sum_{k=1}^q g_{jk}(V_\mathrm{m}^j - V_\mathrm{m}^k)\right) &&\text{(on $\mathcal{R}, \ j=1, \ldots, q$)}, \nonumber\\
\frac{\partial \boldsymbol{X}^j}{\partial t} = \boldsymbol{G}^j(V_\mathrm{m}^j,\boldsymbol{X}^j) &&\text{(on $\mathcal{R}, \ j=1, \ldots, q$)}\label{eq: multidomain auxiliary}.
\end{align}

\subsection{Numerical Methods}\label{numerical methods}

In this section, we discuss our numerical methods to solve the PDE system.
For the purpose of clarity and to avoid any repetition, the discussion here will be focused on the bi-domain equations.
The method can be naturally extended to the multi-domain equation.

\subsubsection{Spatial Discretization}\label{spatial discretization}

We discretize our system in space using non-uniform finite differences.
We use a spherical coordinate system $(r,\theta,\varphi)$, where $r$ is the distance of a point to the origin, $\theta$ is the polar angle in the $xy$-plane, and $\varphi$ is the signed latitude from the $xy$-plane.

\paragraph{Nonuniform Tensor Product Grid}

For our finite difference discretization we opt to use a nonuniform tensor product grid.	
We choose a nonuniform grid as the retina, being an active domain which receives a variety of light stimuli, requires a very fine grid.
Extending such a grid to the rest of the eye, noting the difference in size, would make the computations unwieldy.
A completely nonuniform grid (i.e. one that is not a tensor product) would be even more efficient, but difficult to implement correctly, especially considering all the boundary conditions in the system.

\paragraph{Discretized PDE System in Spherical Coordinates}

The gradient in spherical coordinates is 
\begin{equation}\label{eq: gradient}
\nabla = \pard{}{r}\bar{r} + \frac{1}{r\cos\varphi}\pard{}{\theta}\bar{\theta}+\frac{1}{r}\pard{}{\varphi}\bar{\varphi},
\end{equation}
in which $\bar{r},\bar{\theta},\bar{\varphi}$ are unit vectors in the radial, polar, and latitudinal directions, respectively, which locally form a spherical basis on $\R^3$ (not including the $z$-axis).
Considering the column packing of the retina, we restricted the types of conductivity tensors, in this spherical basis, to be of the form
\begin{align*}
\mu_{\mathrm{x}} &= \begin{bmatrix}
					\mu_\mathrm{x}^r &0	&0 \\
					0	&\mu_\mathrm{x}^\theta	&0\\
					0	&0	&\mu_\mathrm{x}^\varphi
				\end{bmatrix}	&\text{for x}=\text{ i, e}
\end{align*}
in which $\mu_\mathrm{x}^r, \mu_\mathrm{x}^\theta, \mu_\mathrm{x}^\varphi$ are the conductivities in the radial, polar, and latitudinal directions, respectively.
Off of the retina, on $\mathcal{S} \setminus \mathcal{R}$, considering we are in the electrically passive vitreous chamber, one can justifiably consider $\mu_s$ to be a nonnegative real-valued function (meaning $\mu_\mathrm{s}^r=\mu_\mathrm{s}^\theta=\mu_\mathrm{s}^\varphi$).
Continuing generically, locally we can write $\mu_\mathrm{x}$ as a linear transformation
\begin{multline}\label{eq: conductivities}
\mu_\mathrm{x}(r,\theta,\varphi)(v) = \mu_\mathrm{x}^r(r,\theta,\varphi) \bar{r}^*(v) \bar{r} \\
+ \mu_\mathrm{x}^\theta(r,\theta,\varphi) \bar{\theta}^*(v) \bar{\theta} + \mu_\mathrm{x}^\varphi(r,\theta,\varphi) \bar{\varphi}^*(v) \bar{\varphi} \hspace{1 cm} \text{for x}=\text{ i, e, s,}
\end{multline}
in which $\bar{r}^*,\bar{\theta}^*,\bar{\varphi}^*$ are the standard linear functionals associated with the basis $\bar{r},\bar{\theta},\bar{\varphi}$.
Using \cref{eq: gradient,eq: conductivities}, we can express \crefrange{eq: bidomain diffusion}{eq: bidomain auxiliary} in spherical coordinates.
For example, \cref{eq: bidomain diffusion} can be written in spherical coordinates as
\begin{align} \label{eq: spherical diffusion}
\pard{\mu_\mathrm{s}^r}{r} \pard{\phi_\mathrm{s}}{r} + \mu_\mathrm{s}^r \pardd{\phi_\mathrm{s}}{r} + 2\frac{\mu_\mathrm{s}^r}{r}\pard{\phi_\mathrm{s}}{r} +&\frac{1}{r^2\cos^2\varphi}\left(\pard{\mu_\mathrm{s}^\theta}{\theta} \pard{\phi_\mathrm{s}}{\theta} + \mu_\mathrm{s}^\theta \pardd{\phi_\mathrm{s}}{\theta} \right)  \nonumber \\ 
-& \frac{\mu_\mathrm{s}^\varphi \sin\varphi}{r^2\cos\varphi}\pard{\phi_\mathrm{s}}{\varphi} +\frac{1}{r^2}\left(\pard{\mu_\mathrm{s}^\varphi}{\varphi} \pard{\phi_\mathrm{s}}{\varphi} + \mu_\mathrm{s}^\varphi \pardd{\phi_\mathrm{s}}{\varphi} \right)= 0.
\end{align}

We use the non-uniform centered finite difference formulas to discretize the first and second derivatives.
Let $\{r^0=0, \ldots, r^n=r_\mathrm{eye}\}$ be the discretization points in the radial direction and $\phi^{(i,j,k)}$ correspond the value of $\phi$ at the $(i,j,k)$ node, where $i,j,k$ are indices for the radial, polar, and latitudinal directions, respectively.
The equations we use are
\begin{align}
\pard{\phi^{(i,j,k)}}{r} = \frac{\phi^{(i+1,j,k)} - \gamma_i^2\phi^{(i-1,j,k)} -(1-\gamma_i^2)\phi^{(i,j,k)}}{(1+\gamma_i)h_{i+1}}, \label{eq: first derivative}\\
\pardd{\phi^{(i,j,k)}}{r} = 2\gamma_i\frac{\phi^{(i+1,j,k)} + \gamma_i\phi^{(i-1,j,k)} -(1+\gamma_i)\phi^{(i,j,k)}}{(1+\gamma_i)h_{i+1}^2}, \label{eq: second derivative}
\end{align} 
in which $h_{i+1} = r^{i+1}-r^i$ and $\gamma_i = \frac{h_{i+1}}{h_i}$.

Equations \cref{eq: spherical diffusion,eq: first derivative,eq: second derivative} (and similar equations for the polar and latitudinal directions) provide a linear relation between the potential values on the grid. From \cref{eq: bidomain diffusion} and boundary conditions \cref{eq: potential continuity,eq: curr continuity}, we have
\begin{equation}\label{eq: discretized diffusion}
A \boldsymbol{\phi_\mathrm{s}} = B\boldsymbol{\phi_\mathrm{e}},
\end{equation}
in which $A$ and $B$ are sparse matrices, and $\boldsymbol{\phi_\mathrm{s}}$ and $\boldsymbol{\phi_\mathrm{e}}$ are vectors containing the grid point values of the corresponding functions stored in reverse lexicographical order.
Similarly, \cref{eq: bidomain conservation} becomes 
\begin{equation}\label{eq: discretized conservation}
C [\boldsymbol{\phi_\mathrm{s}}; \boldsymbol{\phi_\mathrm{e}}; \boldsymbol{V_\mathrm{m}}] = \boldsymbol{0},
\end{equation}
in which $C$ is also a sparse matrix, $\boldsymbol{V_\mathrm{m}}$ is a vector containing the grid point values of $V_\mathrm{m}$ stored in reverse lexicographical order, and the square bracket notation $[ \ \cdot \ ; \ \cdot \ ]$ denotes concatenating two or more vectors.
Unambiguously, given two vectors $\boldsymbol{x} = (x_1,\ldots,x_n), \boldsymbol{y} = (y_1, \ldots, y_m)$, using the square bracket notation we get $[\boldsymbol{x}; \boldsymbol{y}] = (x_1,\ldots, x_n,y_1,\dots,y_m)$.
\Cref{eq: bidomain voltage,eq: bidomain auxiliary} become a system of ODEs for the values of $V_\mathrm{m}$ and $\boldsymbol{X}$ on the grid,
\begin{equation}\label{eq: discretized voltage and auxiliary}
\frac{d [\boldsymbol{V_\mathrm{m}};\boldsymbol{X}]}{dt} = \boldsymbol{G}(\boldsymbol{\phi_\mathrm{s}},\boldsymbol{\phi_\mathrm{e}},\boldsymbol{V_\mathrm{m}},\boldsymbol{X}).
\end{equation}
Note that we abuse the notation by writing $\boldsymbol{X}$ for both the function and its values on the grid.
We also note that, in this setup, $\boldsymbol{G}$ is a nonlinear function as a result of the nonlinearity in the transmembrane currents, $I_{\mathrm{m}}$.
\Cref{eq: discretized diffusion,eq: discretized conservation,eq: discretized voltage and auxiliary} are a system of differential algebraic equations (DAEs)~\cite{Petzold}.

\paragraph{Boundary Conditions}

To setup the detailed discussion of the boundary conditions , let $\{r^0>0, \ldots, r^i = r_\text{eye}-r_\mathrm{retina}, \ldots, r^n = r_\mathrm{eye}\}$ be the grid points in the radial direction.
We do not include $r=0$ in our grid as it is a coordinate singularity.
When the finite differences require the value at the origin, it is given by taking the average values of grid points on the sphere of radius $r^0$.

The potential $\phi_\mathrm{e}$ appears in \cref{eq: discretized diffusion} as a result of the retinal boundary conditions imposed on the system. In fact, the discretization resulting in \crefrange{eq: discretized diffusion}{eq: discretized voltage and auxiliary} incorporates the boundary conditions.
The Neumann boundary conditions in our system are
\begin{align}
n_\mathrm{x} \cdot (\mu_{\mathrm{i}} \nabla V_\mathrm{m}) &= -n_\mathrm{x} \cdot (\mu_{\mathrm{i}} \nabla \phi_{\mathrm{e}})	& \text{(on $\partial\mathcal{R}_\mathrm{x}$ for $\mathrm{x} = \mathrm{o}, \mathrm{i}, \ell$),} \label{eq: boundary intracellular}\\
n_\mathrm{x} \cdot (\mu_{\mathrm{e}} \nabla \phi_{\mathrm{e}}) &= n_\mathrm{x} \cdot (\mu_{\mathrm{s}} \nabla \phi_{\mathrm{s}}) & \text{(on $\partial\mathcal{R}_\mathrm{x}$ for $\mathrm{x} = \mathrm{i}, \ell$),} \label{eq: current continuity}\\
n_\mathrm{o} \cdot (\mu_{\mathrm{e}} \nabla \phi_{\mathrm{e}}) &= 0	& \text{(on $\partial\mathcal{R}_\mathrm{o}$),} \label{eq: boundary extracellular}\\
n_\mathrm{s} \cdot (\mu_{\mathrm{s}} \nabla \phi_{\mathrm{s}}) &= 0 & \text{(on $\partial\mathcal{S} \setminus \partial\mathcal{R}_\mathrm{o}$),} \label{eq: boundary surface}
\end{align}
which were obtained from the original boundary conditions by using $\phi_\text{i} = V_\mathrm{m} + \phi_\mathrm{e}$, to eliminate $\phi_\text{i}$.

The conditions in \cref{eq: boundary extracellular,eq: boundary surface} are tantamount to $\pard{\phi}{r}=0$ (we dropped the subscript as the treatment of \cref{eq: boundary extracellular,eq: boundary surface} are similar).
To impose this condition we use the grid extension method~\cite{LeVeque} and compute the radial derivative according to the following equations
\begin{align*}
\pard{\phi^{(n,j,k)}}{r} &= 0, \\
\pardd{\phi^{(n,j,k)}}{r} &= 2\gamma_n\frac{\phi^{(n+1,j,k)} + \gamma_n\phi^{(n-1,j,k)} -(1+\gamma_n)\phi^{(n,j,k)}}{(1+\gamma_n)h_{n+1}^2} = \frac{2\phi^{(n-1,j,k)} -2\phi^{(n,j,k)}}{h_{n}^2},
\end{align*}
in which $\gamma_n =1$ as we choose $h_{n+1} = h_n$.

Now we deal with the first group of boundary conditions, and we take \cref{eq: current continuity}, with x = i, as our working example.
After expansion, \cref{eq: current continuity} becomes $\mu_\mathrm{e}^r \pard{\phi_e}{r}= \mu_\mathrm{s}^r\pard{\phi_s}{r}$ (on $\partial R_i$).
Using the one sided finite difference formula on $\phi_\mathrm{s}$ and $\phi_\mathrm{e}$ we get
\begin{equation*}
\mu_\mathrm{e}^r\frac{\phi_\mathrm{e}^{(i,j,k)}-\phi_\mathrm{e}^{(i-1,j,k)}}{h_i} = \mu_\mathrm{s}^r\frac{\phi_\mathrm{s}^{(i,j,k)}-\phi_\mathrm{s}^{(i-1,j,k)}}{h_i}, 
\end{equation*}
which can be solved to give
\begin{equation}
\phi_\mathrm{e}^{(i-1,j,k)} = \frac{\mu_\mathrm{s}^r}{\mu_\mathrm{e}^r}\phi_\mathrm{s}^{(i-1,j,k)} + \left(1-\frac{\mu_\mathrm{s}^r}{\mu_\mathrm{e}^r}\right)\phi_\mathrm{e}^{(i,j,k)}. \label{eq: ghost point}
\end{equation}
Again using the grid extension method, \cref{eq: ghost point} can be used in \cref{eq: first derivative,eq: second derivative} to compute the derivative on $\mathcal{R}_\mathrm{i}$, which finishes our treatment of the boundary conditions.

\subsubsection{Stepping Through Time Using an Adaptive Time-stepper}\label{time-stepping}
After spatial discretization, we must solve the DAE in time.
This involves solving an equation of the form $\frac{d\boldsymbol{x}}{dt} = \boldsymbol{f}(\boldsymbol{s},t,\boldsymbol{x})$.
We can assume that we are interested in $\frac{d\boldsymbol{x}}{dt} = \boldsymbol{f}(t,\boldsymbol{x})$ instead, since we can solve for the potentials from the membrane voltage using \cref{eq: discretized diffusion,eq: discretized conservation}.
To solve the differential equation, we use the second order, variable step-size, backward differentiation formula (BDF2).
That is, to solve for the value of $\boldsymbol{x}$ at time $t_n$ (i.e. $\boldsymbol{x}_n$) we must solve the following equation
\[
	\boldsymbol{x}_n - c_{1}\boldsymbol{x}_{n-1} - c_{2} \boldsymbol{x}_{n-2} = c_0 \beta \boldsymbol{f}(t_n,\boldsymbol{x}_n),
\]
with parameters from the BDF2 scheme \cref{eq: BDF2}.
We do so using Newton's method to find the root of the function $\boldsymbol{F}(\boldsymbol{x}) = \boldsymbol{x} - c_1\boldsymbol{x}_{n-1} - c_{2} \boldsymbol{x}_{n-2} - c_0 \beta \boldsymbol{f}(t_n,\boldsymbol{x})$.
So the equation we are interested in solving is $\boldsymbol{F}'(\boldsymbol{x}^i) {\Delta \boldsymbol{x}}^{i+1} = -\boldsymbol{F}(\boldsymbol{x}^i)$, in which $\boldsymbol{x}^i$ is solution of the previous Newton iteration and ${\Delta \boldsymbol{x}}^{i+1} = \boldsymbol{x}^{i+1}-\boldsymbol{x}^i$.
This equation and the algebraic constraints can be written as
\[
D^i {\Delta \boldsymbol{x}}^{i+1} = \boldsymbol{r}^i,
\]
in which
\[
\boldsymbol{r^i} = \begin{bmatrix}
\boldsymbol{0} \\
\boldsymbol{0} \\
-\boldsymbol{F}(\boldsymbol{x}^i) \\
0
\end{bmatrix}
\]
and $D^i$ is an $(n+1)\times n$ matrix composed of an arrangement of $A, B, C$, and the Jacobian $\boldsymbol{F}'(\boldsymbol{x}^i)$.
The extra row at the bottom of $D^i$ and $\boldsymbol{r}^i$ is to impose \cref{grounding at infinity}.
Solving this Jacobian update equation is the most time consuming step in the simulation (see \cref{Jacobian update}).

It is worth noting that when light hits the retina and activates the opsin proteins, it triggers a cascade of events bringing about rapid changes in the retina.
Once the light stimulus is gone, retinal cells return to a resting state somewhat slowly.
The presence of these multiple time scales in our model and the costly computation of each time-step, necessitates a variable time-step solver.
The complexity of the system we are studying, as well as all the possible variability in light stimuli make it clear that we must employ an adaptive time-stepping method as opposed to a variable time-stepping method with preset time-step values.

\paragraph{General Scheme and Underlying Numerical Method}\label{adaptive time-stepping}
We are using this time-stepper to solve the differential equation in our DAE system, namely
\begin{align*}
&\frac{d [\boldsymbol{V_\mathrm{m}};\boldsymbol{X}]}{dt} = \boldsymbol{G}(\boldsymbol{\phi_\mathrm{s}},\boldsymbol{\phi_\mathrm{e}},\boldsymbol{V_\mathrm{m}},\boldsymbol{X})& \text{(on $\mathcal{R}$)}.
\end{align*}
To simplify notation, the details of the time-stepper will be explained as to solve the generic equation $\frac{d\boldsymbol{x}}{dt} = \boldsymbol{f}(t,\boldsymbol{x})$.
Let $\boldsymbol{x}_n$ denote the computed approximation of $\boldsymbol{x}(t_n)$ and $h_n = t_{n+1}-t_n$ be the step-size of the $n^{\text{th}}$ step.

The basic idea of the method is a coarse-fine computation, which has been well studied and is regularly used to study various physical systems~\cite{Hairer, Pugh}.
Starting at $\boldsymbol{x}_n$ we compute our coarse approximation, $\boldsymbol{x}_{n+1}^\mathrm{c}$, using one step of size $h_n$.
We then go back to $\boldsymbol{x}_n$ and compute the fine approximation, $\boldsymbol{x}_{n+1}^\mathrm{f}$, using two steps of size $h_n/2$.
Finally, we use both these approximations to estimate our coarse local truncation error, $\epsilon_c$, (see \cref{error estimate}) and if the error is suitable the step-size, $h_n$, and the coarse approximation are accepted.
If the error is too small or too large then a new step-size is chosen according the formula
\[
h_n \cdot\min\left\{\max\left\{\left(\frac{tol}{\epsilon_c}\right)^\frac{1}{p},\eta_\mathrm{min} \right\},\eta_\mathrm{max}\right\},
\]
in which $tol$ is the desired local truncation error (LTE), $p$ is the order of the LTE (calculated in \cref{error estimate}), and $\eta_\mathrm{max},\eta_\mathrm{min}$ are safety factors that prevent $h_n$ from drastically changing from one iteration to the next.
We reject any time-steps with an error estimate outside the desired range.
Once a new $h_n$ is chosen, we recompute $\boldsymbol{x}_{n+1}^\mathrm{c}$ and $\boldsymbol{x}_{n+1}^\mathrm{f}$ and repeat the procedure to get a LTE within the accepted range.

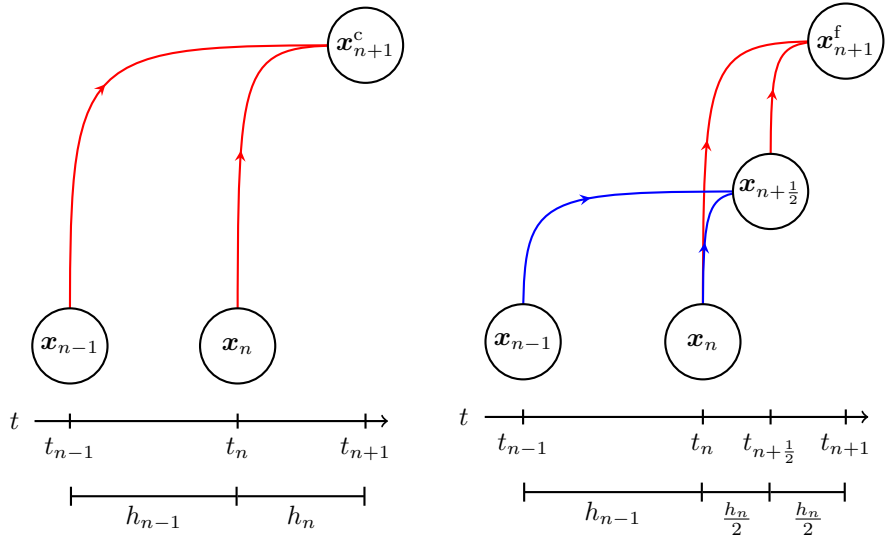
\begin{figure}
\begin{minipage}{0.45\textwidth}
\begin{tikzpicture}[decoration={
	markings,
	mark=at position 0.5 with {\arrow{stealth}}}
    ] 
\pgfmathsetmacro{\x}{0.47}
\pgfmathsetmacro{\y}{2.7}
\pgfmathsetmacro{\z}{4.4}
\pgfmathsetmacro{\tickwidth}{0.1}
\pgfmathsetmacro{\h}{1}
\pgfmathsetmacro{\hh}{5}
\draw[black, thick,->] (0,0) -- (4.7,0) node[left] at (-0.07,0) {$t$};
\foreach \t in {\x,\y,\z} {
	\draw[black,thick](\t,-\tickwidth) -- (\t,\tickwidth);
	}
\draw node[below] at (\x,-\tickwidth) {$t_{n-1}$};
\draw node[below] at (\y,-\tickwidth) {$t_{n}$};
\draw node[below] at (\z,-\tickwidth) {$t_{n+1}$};
\draw[black, thick,|-|] (\x,-1) -- (\y,-1) node[midway,below] {$h_{n-1}$};
\draw[black, thick,-|] (\y,-1) -- (\z,-1) node[midway,below] {$h_{n}$};

\draw[thick,red,postaction={decorate}] (\x,\h) .. controls (\x,\hh) .. (\z,\hh);
\draw[thick,red,postaction={decorate}] (\y,\h) .. controls (\y,\hh) .. (\z,\hh);

\draw[thick,fill=white] (\x,\h) circle(0.5) node at (\x,\h) {$\boldsymbol{x}_{n-1}$};
\draw[thick,fill=white] (\y,\h) circle(0.5) node at (\y,\h) {$\boldsymbol{x}_n$};
\draw[thick,fill=white] (\z,\hh) circle(0.5) node at (\z,\hh) {$\boldsymbol{x}_{n+1}^\mathrm{c}$};
\end{tikzpicture}
\end{minipage}
\begin{minipage}{0.5\textwidth}
\begin{tikzpicture}[decoration={
	markings,
	mark=at position 0.5 with {\arrow{stealth}}}
    ] 
\pgfmathsetmacro{\x}{0.51}
\pgfmathsetmacro{\y}{2.9} 
\pgfmathsetmacro{\z}{3.8} 
\pgfmathsetmacro{\w}{4.8}
\pgfmathsetmacro{\tickwidth}{0.1}
\pgfmathsetmacro{\h}{1}
\pgfmathsetmacro{\hh}{3}
\pgfmathsetmacro{\hhh}{5}
\draw[black, thick,->] (0,0) -- (5.1,0) node[left] at (-0.08,0) {$t$};
\foreach \t in {\x,\y,\z,\w} {
	\draw[black,thick](\t,-\tickwidth) -- (\t,\tickwidth);
	}
\draw node[below] at (\x,-\tickwidth) {$t_{n-1}$};
\draw node[below] at (\y,-\tickwidth) {$t_{n}$};
\draw node[below] at (\z,-\tickwidth) {$t_{n+\frac{1}{2}}$};
\draw node[below] at (\w,-\tickwidth) {$t_{n+1}$};
\draw[black, thick,|-|] (\x,-1) -- (\y,-1) node[midway,below] {$h_{n-1}$};
\draw[black, thick,-|] (\y,-1) -- (\z,-1) node[midway,below] {$\frac{h_{n}}{2}$};
\draw[black, thick,-|] (\z,-1) -- (\w,-1) node[midway,below] {$\frac{h_{n}}{2}$};

\draw[thick,red,postaction={decorate}] (\y,\h) .. controls (\y,\hhh) .. (\w,\hhh);
\draw[thick,red,postaction={decorate}] (\z,\hh) .. controls (\z,\hhh) .. (\w,\hhh);
\draw[thick,blue,postaction={decorate}] (\x,\h) .. controls (\x,\hh) .. (\z,\hh);
\draw[thick,blue,postaction={decorate}] (\y,\h) .. controls (\y,\hh) .. (\z,\hh);

\draw[thick,fill=white] (\x,\h) circle(0.5) node at (\x,\h) {$\boldsymbol{x}_{n-1}$};
\draw[thick,fill=white] (\y,\h) circle(0.5) node at (\y,\h) {$\boldsymbol{x}_n$};
\draw[thick,fill=white] (\w,\hhh) circle(0.5) node at (\w,\hhh) {$\boldsymbol{x}_{n+1}^\mathrm{f}$};
\draw[thick,fill=white] (\z,\hh) circle(0.5) node at (\z,\hh) {$\boldsymbol{x}_{n+\frac{1}{2}}$};

\end{tikzpicture}
\end{minipage}
\caption{Left: The coarse approximation of $\boldsymbol{x}$ at $t_{n+1}$, $\boldsymbol{x}_{n+1}^\mathrm{c}$, is computed using two accepted values, $\boldsymbol{x}_n$ and $\boldsymbol{x}_{n-1}$, which were precomputed at previous time-steps. Right: The fine approximation of $\boldsymbol{x}$ at $t_{n+1}$, $\boldsymbol{x}_{n+1}^\mathrm{f}$, is computed in two steps. First we compute the intermediate value $\boldsymbol{x}_{n+\frac{1}{2}}$, which approximates $\boldsymbol{x}$ at $t_{n+\frac{1}{2}}=t_n + \frac{h_n}{2}$ (shown in blue). Then we compute $\boldsymbol{x}_{n+1}^\mathrm{f}$ using $\boldsymbol{x}_n$ and $\boldsymbol{x}_{n+\frac{1}{2}}$ (shown in red).}
\label{scheme}
\end{figure}

The underlying numerical method used here is the variable step size BDF2 given by
\begin{equation} \label{eq: BDF2}
\boldsymbol{x}_{n+1} - \frac{(1+\omega_n)^2}{1+2\omega_n}\boldsymbol{x}_n + \frac{\omega_n^2}{1+2\omega_n}\boldsymbol{x}_{n-1} = h_n \frac{1+\omega_n}{1+2\omega_n}\boldsymbol{f}(t_{n+1},\boldsymbol{x}_{n+1}),
\end{equation}
in which $\omega_n = \frac{h_n}{h_{n-1}}$.
To compute $\boldsymbol{x}_{n+1}^\mathrm{c}$ we use $\boldsymbol{x}_n$ and $\boldsymbol{x}_{n-1}$.
We also use $\boldsymbol{x}_n$ and $\boldsymbol{x}_{n-1}$ to approximate $\boldsymbol{x}$ at the half-step ($t=t_n + \frac{h_n}{2}$), $\boldsymbol{x}_{n+\frac{1}{2}}$.
We subsequently use $\boldsymbol{x}_{n+\frac{1}{2}}$ and $\boldsymbol{x}_n$ to compute $\boldsymbol{x}_{n+1}^\mathrm{f}$.
A schematic sketch of the method is provided in \cref{scheme}.
It should be clear from \cref{scheme} that the coarse and fine computation are completely independent and could be carried out in parallel to enhance performance.
However, since there are only two parallel tasks of modest duration, any performance gains will be diminished by the overhead costs of parallelization.

\paragraph{Local Truncation Error Estimate} \label{error estimate}
Let us approximate $\mathbf{LTE} = \boldsymbol{x}(t_{n+1}) - \boldsymbol{x}_{n+1}$ for our method.
Using \labelcref{eq: BDF2}, $\boldsymbol{f}(t_{n+1},\boldsymbol{x}_{n+1})$ $\approx \boldsymbol{x}'(t_{n+1})$, and a Taylor series about $t_n$ (assuming $\boldsymbol{x}_n = \boldsymbol{x}(t_n)$ and $\boldsymbol{x}_{n-1} = \boldsymbol{x}(t_{n-1})$) we get
\begin{equation}\label{eq: LTE}
\mathbf{LTE} = -\frac{1+\omega_n}{1+2\omega_n} \frac{\boldsymbol{x}'''(t_n)}{3!}h_n^2(h_n+h_{n-1}) + O(h^4)
\end{equation}
in which $h^4$ is understood as a product of $h_n$ and $h_{n-1}$ with combined powers of $4$.
We now compute the fine LTE, $\boldsymbol{\epsilon_\mathrm{f}} = \boldsymbol{x}(t_{n+1}) - \boldsymbol{x}_{n+1}^\mathrm{f}$ (coarse LTE is obtained directly from \labelcref{eq: LTE}).
Using \labelcref{eq: BDF2} with equal step-sizes $\frac{h_n}{2}$ on the second fine step we get
\[
\boldsymbol{\epsilon_\mathrm{f}} = \boldsymbol{x}(t_{n+1}) - \left(\frac{4}{3}\boldsymbol{x}_{n+\frac{1}{2}} - \frac{1}{3}\boldsymbol{x}_{n} + \frac{h_n}{2} \frac{2}{3}\boldsymbol{f}(t_{n+1},\boldsymbol{x}_{n+1})\right).
\]
We then use \labelcref{eq: BDF2} a second time but with step-size $h_{n-1}$ and $\frac{h_n}{2}$ on $\boldsymbol{x}_{n+\frac{1}{2}}$ and obtain the fine LTE to be
\[
\boldsymbol{\epsilon_\mathrm{f}} = -\frac{1}{3! \cdot 3} \ \frac{1}{1+\omega_n} \boldsymbol{x}'''(t_n) h_n^2 h_{n-1} - \frac{1}{3! \cdot 2} \ \frac{1+\frac{\omega_n}{2}}{1+\omega_n} \boldsymbol{x}'''(t_n) h_n^3 + O(h^4).
\]
The third derivative in \cref{eq: LTE} can be approximated using our coarse-fine approach since
\begin{multline}
\boldsymbol{x}_{n+1}^\mathrm{c} - \boldsymbol{x}_{n+1}^\mathrm{f} = \boldsymbol{\epsilon_\mathrm{f}} - \boldsymbol{\epsilon_\mathrm{c}} \approx \left(\frac{\omega_n^2 + \frac{7}{4}\omega_n + \frac{1}{2}}{(1+2\omega_n)(1+\omega_n)}\frac{\boldsymbol{x}'''(t_n)}{3!}\right)h_n^3 \\
+ \left(\frac{\omega_n^2+ 2\omega_n + \frac{2}{3}}{(1+2\omega_n)(1+\omega_n)}\frac{\boldsymbol{x}'''(t_n)}{3!}\right)h_n^2h_{n-1}.
\end{multline}
Solving for $\boldsymbol{x}'''(t_n)$ and substituting into $\boldsymbol{\epsilon_\text{c}}$ we get 
\[
\boldsymbol{\epsilon_\text{c}} \approx -\frac{(1+\omega_n)^3}{\omega_n^3+\frac{11}{4}\omega_n^2+\frac{5}{2}\omega_n+\frac{2}{3}}(\boldsymbol{x}_{n+1}^\mathrm{c} - \boldsymbol{x}_{n+1}^\mathrm{f})
\]
and so we are able to estimate the LTE of the coarse step without computing any derivatives using the coarse and fine approximations.
We note that in this case $\boldsymbol{\epsilon_\mathrm{c}}$ is a vector and so we use the max norm, $\epsilon_\mathrm{c} = \norm{\boldsymbol{\epsilon_\mathrm{c}}}_\mathrm{max}$, to determine whether the error lies within the acceptable range or not.

\paragraph{Richardson Extrapolation}
If $\epsilon_\mathrm{c}$ is within the acceptable range, then we can accept the coarse approximation $\boldsymbol{x}_{n+1}^\text{c}$ as the next value.
However, since we already have an approximation for $\boldsymbol{\epsilon_\mathrm{c}} = \boldsymbol{x}(t_{n+1})-\boldsymbol{x}_{n+1}^\text{c}$, we can use this to obtain a numerical scheme of one order higher ($3$ as opposed to $2$ for us) by taking the following linear combination of $\boldsymbol{x}_{n+1}^\text{c}$ and $\boldsymbol{x}_{n+1}^\text{f}$
\[
\boldsymbol{x}(t_{n+1}) = \boldsymbol{x}_{n+1}^\text{c} + \boldsymbol{\epsilon_\mathrm{c}} \approx -\frac{\frac{1}{4}\omega_n^2+\frac{1}{2}\omega_n+\frac{1}{3}}{\omega_n^3+\frac{11}{4}\omega_n^2+\frac{5}{2}\omega_n+\frac{2}{3}}\boldsymbol{x}_{n+1}^\text{c} +\frac{(1+\omega_n)^3}{\omega_n^3+\frac{11}{4}\omega_n^2+\frac{5}{2}\omega_n+\frac{2}{3}} \boldsymbol{x}_{n+1}^\text{f}.
\]
Richardson Extrapolation might not always be best to use due to stability issues~\cite{Pugh}.

\subsubsection{Solving the Linear Jacobian Update Equation}\label{Jacobian update}
As discussed in \cref{time-stepping}, the linear Jacobian update equation can be written as
\begin{align}\label{eq: Jacobian update}
D^i {\Delta \boldsymbol{x}}^{i+1} = \boldsymbol{r}^i.
\end{align}
After ensuring uniqueness of solution (see \cref{uniqueness}), $D^i$'s are $(n+1) \cross n$, large, sparse matrices.
Solving these updates directly proved to be onerous and expensive.
Hence, we solve \cref{eq: Jacobian update} using a semi-explicit iterative method given by
\begin{align}\label{D'Jakonov}
D^0 {\Delta \boldsymbol{x}}_{j+1}^{i+1} = \boldsymbol{r}_i - (D^i - D^0){\Delta \boldsymbol{x}}_{j}^{i+1},
\end{align}
in which ${\Delta \boldsymbol{x}}_{j}^{i+1}$ is a sequence that converges to ${\Delta \boldsymbol{x}}^{i+1}$ (we choose ${\Delta \boldsymbol{x}}_{0}^{i+1}=\boldsymbol{0}$).
\Cref{D'Jakonov} can be casted as a fixed point iteration based on the splitting of $D^i = D^0 + (D^i - D^0)$~\cite{Saad}.
Historically, this inner iteration is based on the Richardson-D'Jakonov iteration~\cite{Richardson, Djakonov1, Djakonov2}.
Using our terminology, D'Jakonov's proposed inner iteration~\cite{Djakonov2} is based on the splitting $\gamma_j D^i = A + (\gamma_j D^i - A)$, in which $\gamma_j>0$ may vary with each inner iteration.
We choose $A = D^0$ and $\gamma_j =1$.
The motivation for the choice of $D^0$ is it allows us to decompose $D^0$ once and use the decomposition to solve all subsequent iterations (for each application of this Newton-iterative method).
This makes solving \cref{D'Jakonov} significantly faster in comparison to direct solves of \cref{eq: Jacobian update}.
A detailed discussion of computational improvements and accuracy of this scheme is carried out in \cref{direct-iterative comparison}.

As is common with iterative solvers, we choose to solve \cref{eq: Jacobian update} only approximately.
Once the relative residual, 
\[
\frac{\norm{D^i{\Delta \boldsymbol{x}}_j^{i+1} - \boldsymbol{r}^i}_\mathrm{max}}{\norm{\boldsymbol{r}^i}_\mathrm{max}},
\] 
decreases below $10^{-6}$, we halt the iterative solver and accept the iterate ${\Delta \boldsymbol{x}}_j^{i+1}$.
Hence, in this setting, we are solving \cref{eq: BDF2} using an inexact Newton method~\cite{Steihaug, Dembo}.
Details of the inner iterations of this inexact Newton method is also included in \cref{direct-iterative comparison}.

\section{Results and Discussion} \label{results and discussion}

In this section we present and discuss the results of a variety of simulations with various experimental set-ups.
Some of these findings were consistent with the literature while others have no experimental counterpart, but ought to have experiments designed to study them.
For complete details of the various simulations see \cref{appendix}.
We also verify the order of our numerical method and discuss details of the adaptive time-stepper and iterative method used.

\subsection{Desensitization and a-Waves}\label{general observations}

\begin{figure}
\begin{center}
\includegraphics{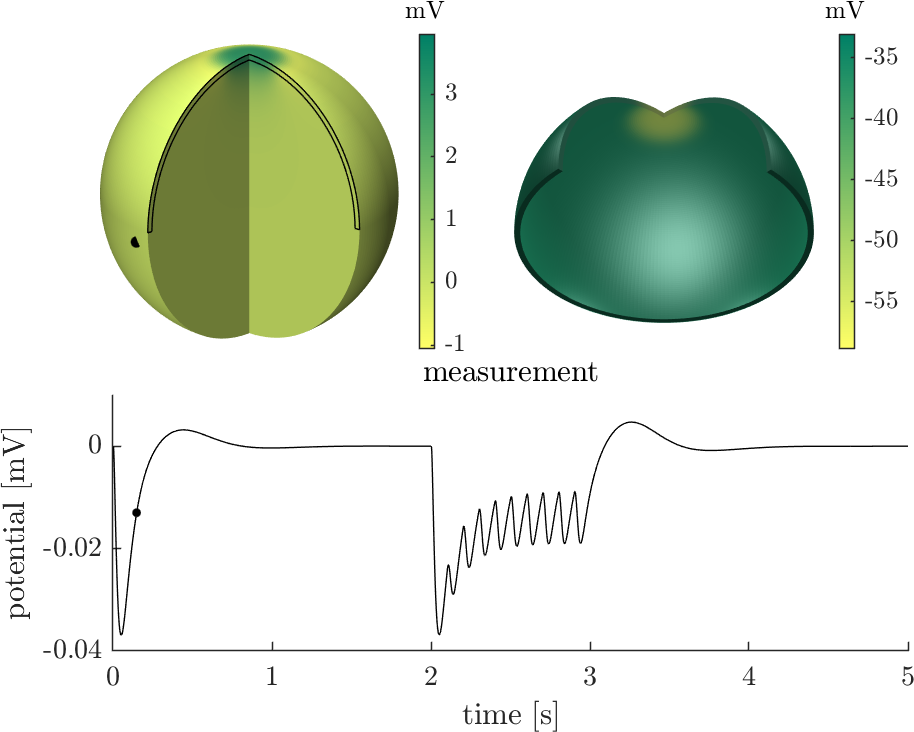}
\end{center}
\caption{Top: (Left) Extracellular potential throughout the eye at $t = 0.15$ s. The region outlined in black is the retina. The black dot is the location of the potential measurement shown in the bottom plot. The effect of the light stimulus is observed in the darker green region near and around the center of the retina. (Right) Voltage on the retina at $t = 0.15$ s. The hyperpolarization, as a result of the light stimulus, is observed in yellow in the central region of the retina. Bottom: Extracellular potential at the specified point (black dot in extracellular potential plot above) on the surface of the eye as a function of time. There are two distinct phases observed in the resulting change from the stimulus, a fast hyperpolarization phase followed by a much slower depolarization phase. See numerical simulation 1 in \cref{appendix} for more details about the set-up of this simulation.}
\label{experiment}
\end{figure}

In numerical simulation 1, the center of the retina is stimulated with spatially Gaussian ($\sigma = 50$), $20$ ms light pulses at $t=0, 2.0, 2.1, 2.2, \ldots, 2.9$ s.
\Cref{experiment} shows the induced change in potential and voltage throughout the surface of the eye and the retina, respectively.
It also includes a plot of the potential as a function of time at a specified location in the eye.
In the time plots of the potential, there are two very different time scales, a fast one occurring right after the light stimulus accompanied by hyperpolarization, and a very slow one in which the cell membrane returns to its depolarized equilibrium.
This property is exactly what we are hoping to exploit in our adaptive time-stepper, since in periods of slow change the solver may take much larger steps.

Many features of retinal physiology are present in our model, some of which are immediately clear from \cref{experiment}. For example, the voltage response between the flashes at time $t=0$ s and $t = 3$ s is similar, while the response is weaker for the subsequent flashes.
This is expected as there was an extended period of time between the first two flashes, while the remaining flashes were temporally close.
This implies that the photoreceptors were able to recover from the first flash but later was desensitized to the rapid light stimuli and so the response was less pronounced.

The a-wave is another feature of retinal physiology that was present in our model.
The two main components of a human ERG reading are the a-wave and the b-wave~\cite{Perlman}.
The a-wave occurs as a result of the hyperpolarization of the photoreceptors~\cite{Perlman}, and hence we were able to observe it in the surface of eye measurements shown in \cref{experiment}.
However, the b-wave occurs as a result of the depolarization of retinal cells that are postsynaptic to photoreceptors~\cite{Perlman}, which are not included in the current model, so we were not able to detect them using our model.
The overshoot observed as the membrane returns to resting potential in \cref{experiment} is different, and much smaller in amplitude, than the typical b-wave~\cite{Perlman}.

\subsection{Calcium Buffering and Delayed Calcium Response in the Inner Segment}

\begin{figure}
\includegraphics{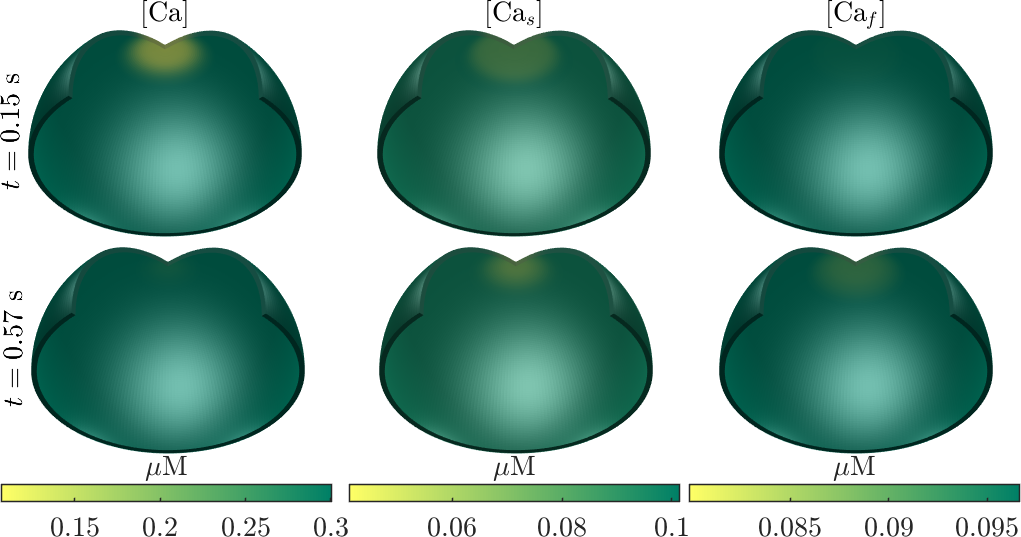}
\caption{Top: The concentration of the outer segment calcium, $[\textrm{Ca}]$, (left), inner segment submembrane calcium, $[\textrm{Ca}_s]$, (center), inner segment central space calcium, $[\textrm{Ca}_f]$, (right) at $t = 0.15$ s. The associated color bar of each plot is located the bottom of the plot's column. The outer segment calcium is quicker to respond to light stimuli than the inner segment calcium. As the movement of calcium is buffered, central space calcium response is not yet visible.  Bottom: similar to the top row but at $t = 0.57$ s. Inner segment calcium response to light stimuli is longer lasting than that of the outer segment. There are no observed differences between submembrane and central space calcium in terms of duration of response.}
\label{calcium}
\end{figure}

The model can be used to study other aspects of the retina as well.
For instance, using the same numerical simulation mentioned above, we were able to study calcium in both the outer segment and the inner segment of the photoreceptors. 
\Cref{calcium} shows the concentrations of the outer segment calcium, $[\textrm{Ca}]$, inner segment submembrane calcium, $[\textrm{Ca}_s]$, and inner segment central space calcium, $[\text{Ca}_f]$ at times $t= 0.15$ s and $t= 0.57$ s.
The distinction between the submembrane calcium and central space calcium reflects research showing that the movement of calcium is controlled in cells using various buffers~\cite{McNaughton}.
Using the model, we observe the time delay between the calcium response to the light stimulus in the outer and inner segment.
This is expected as the outer segment is the cite of photon absorption, which marks the beginning of phototransduction~\cite{Dowling}.
Furthermore, we observe the delay in the response between the submembrane and central space calcium due to the buffers action on calcium.
We also observe that the response in the outer segment is longer lasting than that of the inner segment, which has yet to be observed experimentally.

\subsection{Gap Junctional Effect on Inner Segment Calcium Concentration}

\begin{figure}
\includegraphics{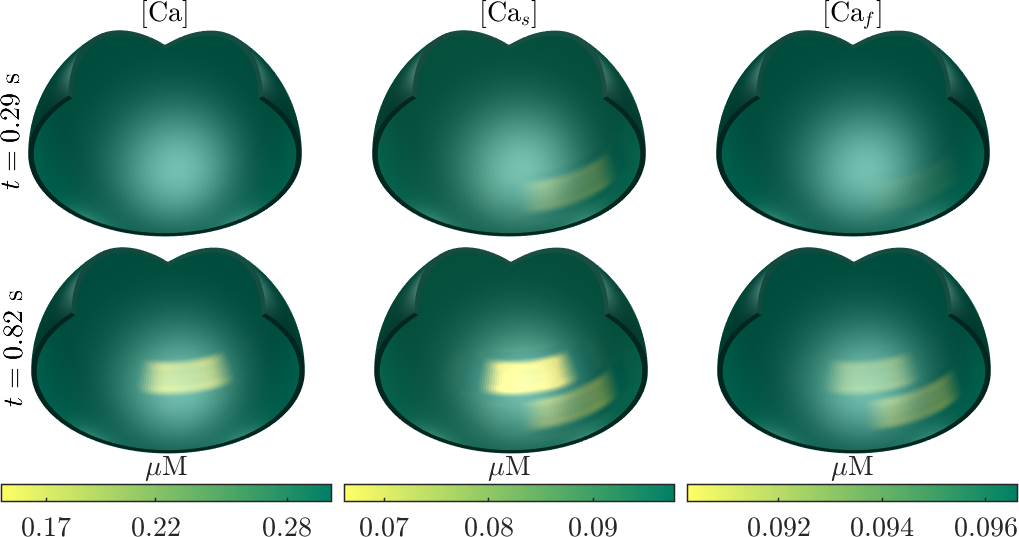}
\caption{Top: The concentration of the outer segment calcium, $[\textrm{Ca}]$, (left), inner segment submembrane calcium, $[\textrm{Ca}_s]$, (center), inner segment central space calcium, $[\textrm{Ca}_f]$, (right) of the L-cones domain at $t = 0.29$ s. The associated color bar of each plot is located the bottom of the plot's column. The inner segment calcium concentration is affected by light stimuli to another photoreceptor. As the movement of calcium is buffered, central space calcium response is not yet clearly visible. No outer segment response is observed. Bottom: similar to the top row but at $t = 0.82$ s. The newly observed response in all three plots is a result of a light stimuli to the L-cones domain, unlike the previous response, whose affects can still be observed in both inner segment calcium concentrations.}
\label{gap junction}
\end{figure}

In numerical simulation 2, we used the multi-domain framework to model four active domains (rods and long (L-), medium (M-), and short wavelength cones (S-cones)).
Each active domain was given a disjoint light stimulus at distinct times (see \cref{appendix}).
The goal of this study was to see whether the various calcium concentrations in an active domain can be affected solely through gap junctions.
Our findings indicate that this is possible for the inner segment calcium concentrations only.
\Cref{gap junction} shows the various calcium concentration of the L-cones domain.
It shows that even in the absence of a light stimuli to the L-cones domain, both the inner segment submembrane calcium concentration and the inner segment central space calcium concentration are affected by the light stimuli to the rods domain.
However, no change was detected in the outer segment calcium concentration.
While we found no experimental evidence supporting these observations, we believe they will hold considering that calcium plays a key role in phototransduction (contributing about a fourth of the photocurrent)~\cite{Torre}.
The observations are also consistent with the findings that gap junctions occur far away from the outer segments, with some occurring in the inner segments~\cite{Raviola,Kolb1,Smith,Cohen,Kolb2}.

\subsection{Convergence Study}\label{convergence study}
Since the underlying time-stepping scheme in our solver is BDF2, we expect to have a global error that is $O(\Delta t^2)$.
We confirm this with a convergence analysis using numerical simulation 3, in which light stimuli is constant in order to preserve $C^3$ requirement shown in \cref{error estimate}.

\begin{figure}
\includegraphics{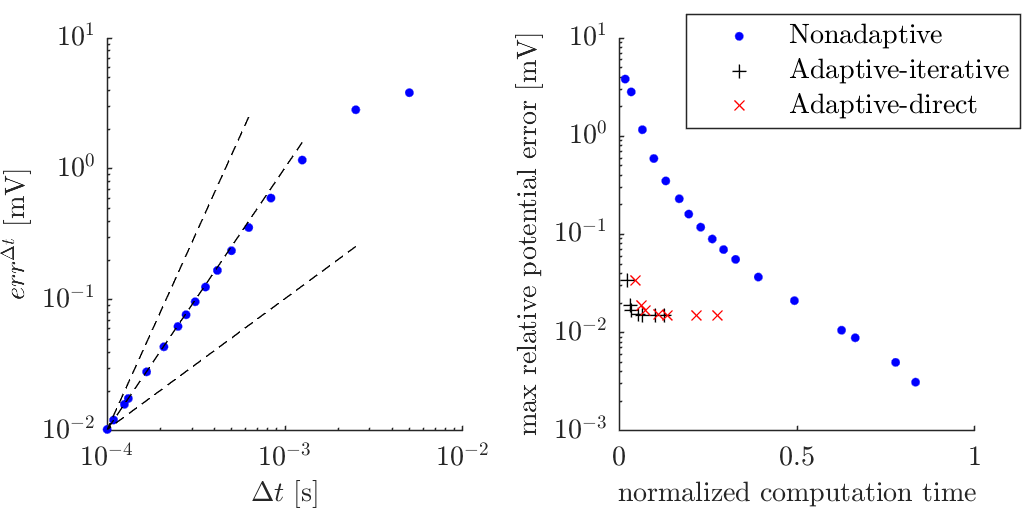}
\caption{Left: A convergence study for our time-stepping method showing the relation between ${err}^{\Delta t}$ and the step-size. The dashed lines have slopes $1$, $2$, and $3$. The expected order of $2$ is observed.
Right: The relation between the wall time and the maximum relative potential error (as compared to the finest constant simulation) for various adaptive-iterative ($\circ$), adaptive-direct ({\color{red}$\cross$}), and constant time-step ({\color{blue}$\bullet$}) simulations. Accuracy and simulation time are, roughly, inversely proportional. In some cases, the adaptive time-stepper decreases the error by a factor of around $100$ as compared to a simulation of similar wall time. The difference between the various adaptive simulations is the $tol$ value chosen (see \cref{adaptive time-stepping}). We ran adaptive simulations (both iterative and direct) for $tol = 5 \cdot 10^{-3}, 10^{-3}, 5 \cdot 10^{-4}, 10^{-4}, 5 \cdot 10^{-5}, 10^{-5}, 5 \cdot 10^{-6}$ (see \cref{fig: direct-iterative comparison})}
\label{convergence and comparison}
\end{figure}

Since the true solution is not available for our numerical simulations, we use the approximation
\begin{equation}\label{relative error}
{err}^{\Delta t} \coloneqq \frac{\norm{\boldsymbol{V_\mathrm{m}}^{\Delta t} - \boldsymbol{V_\mathrm{m}}^{\Delta t^*}}_\text{max}}{1-\left(\frac{\Delta t^*}{\Delta t}\right)^p} \approx C{\Delta t}^p
\end{equation}
in which $\boldsymbol{V_\mathrm{m}}^{\Delta t}$ is the solution $\boldsymbol{V_\mathrm{m}}$ using constant step-size $\Delta t$, $\Delta t^*$ is the finest constant step-size used to approximate the solution to numerical simulation 3, $C$ is a constant, and $p$ is the order of the method.
The modification term $1-\left(\frac{\Delta t^*}{\Delta t}\right)^p$ approaches unity as ${\Delta t}^*$ vanishes since $\boldsymbol{V_\mathrm{m}}^{\Delta t^*}$ becomes the true value, yielding the usual error approximation formula.
\Cref{convergence and comparison} shows the logarithmic relation between ${err}^{\Delta t}$ error and $\Delta t$.
We obtain slope $p=2$ confirming the expected order of our method.

\subsection{Wall Time and Accuracy}\label{nonadaptive-adaptive comparison}
We also use numerical simulation 3 to compare the wall time and accuracy of the adaptive time-stepper and the constant time-stepper.
\Cref{convergence and comparison} shows that the adaptive time-stepper, for most choices of $tol$ values (see \cref{adaptive time-stepping}), improves the accuracy by a factor of at least $10$, as compared to the constant time-stepper with similar wall time.
While the adaptive time-stepper is neither the most accurate nor the fastest simulation, it seems to be the best compromise.

This result depends on the type of experiment conducted.
For example, an experiment with a shorter light pulse would increase the efficiency of the adaptive time-stepper as it will use larger time-steps in the absence of light stimuli (see \cref{adaptive fine details}).
On the other hand, an experiment with multiple light pulses would decrease the efficiency of the adaptive time-stepper as it would need to adjust the step-size numerous times.
Such light stimuli would also affect the constant time-stepper as only small step-sizes would yield reasonably accurate result. 

\subsection{In-depth Analysis of Adaptive Time-stepper}\label{adaptive fine details}

\begin{figure}
\includegraphics{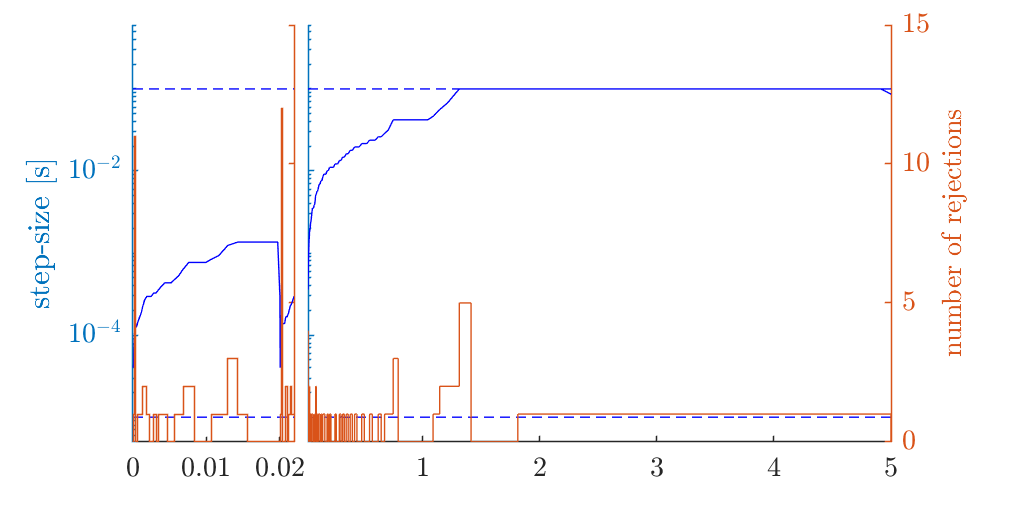}
\caption{Left: Step-size as a function of time. The dashed lines are the minimum and maximum allowed step-sizes, ${\Delta t}_{\text{min}}$ and ${\Delta t}_\text{max}$. Around $t=0.02$ s the step-size decreases sharply as we force the time-stepper to step to certain critical times such as the beginning and end of light stimuli. As the dynamics of the photoreceptors change less frequently (due to absence of any light stimuli) the adaptive-stepper uses larger and larger time-steps. Towards the end of the simulation, the decision of the adaptive-stepper to increase the step-size (as indicated by constant single rejections) is overridden by a safety feature to keep the step-size below ${\Delta t}_\text{max}$.}
\label{fig: adaptive fine details}
\end{figure}

We use numerical simulation 4 to study the adaptive timer-stepper in greater detail.
\cref{fig: adaptive fine details} offers a deeper look on how the adaptive time-stepper works.
It shows how the step-size gradually increases until a suitable step-size is found.
This suitable step-size varies depending on the situation, for example, in the presence of a light stimulus ($t \leq 0.02$ s), the step-size that the adaptive time-stepper stabilized on was much smaller than that in the absence of light ($t > 0.02$).
This is expected as the photoreceptor dynamics are undergoing much more rapid change in the presence of light.
Another feature shown in \cref{fig: adaptive fine details} is the gradual, rather than sudden, change of step-size.
This is evident as for most time-steps at most one rejection was made.

\cref{fig: adaptive fine details} also shows a few safety guards put in place for the adaptive time-stepper.
These include a maximum and minimum allowed step-size, and forcing the time-stepper to step to certain, critical, time values.
The reason for the latter is that when light stimulus shuts off we lose differentiability and so using BDF2 would not be justified.
Instead, we step to the critical time and use improved forward Euler, whose order is the same as BDF2, to take the subsequent first step.

\subsection{Direct and Iterative Solver Comparison}\label{direct-iterative comparison}

\begin{figure}
\includegraphics{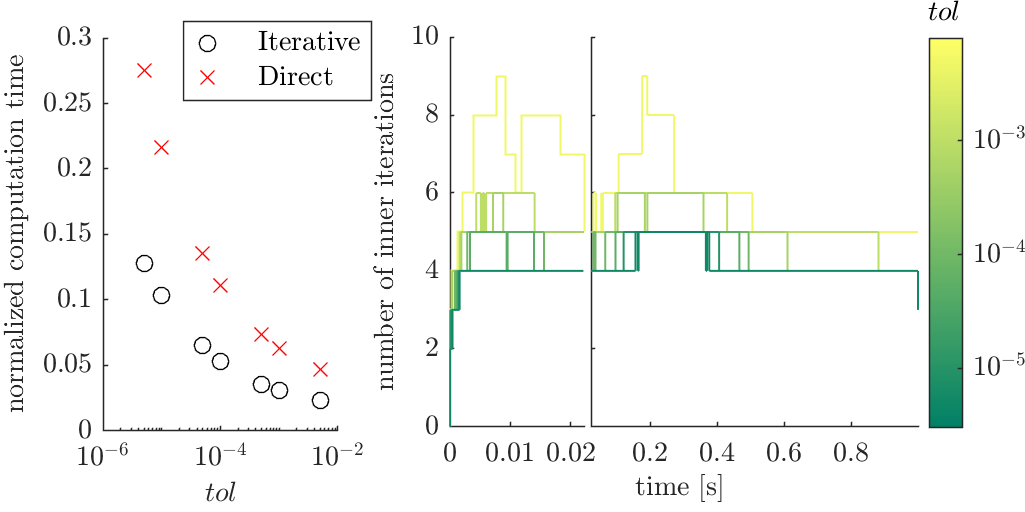}
\caption{Left: Normalized computation time of the adaptive-iterative and adaptive-direct solvers as a function of $tol$ values (see \cref{adaptive time-stepping}). For all $tol$ values tested, the iterative solver improves the wall time of simulations by roughly a factor of $2$ while retaining the same accuracy level (as shown in \cref{convergence and comparison}).
Right: The number of inner iterations required (for all $tol$ values tested) as a function of time. The colorbar on the right indicates the $tol$ values. For all simulations the maximum number of inner iterations needed is less than $10$. We observe that $tol$ values and number of inner iterations required have a positive correlation. This is expected as simulations with lower $tol$ values take smaller step-sizes and so the solution to the Jacobian update equation \cref{eq: Jacobian update} is closer to the initial guess as compared to those with higher $tol$ values.}
\label{fig: direct-iterative comparison}
\end{figure}

In \cref{Jacobian update}, we proposed solving the Jacobian update equation iteratively, using a Richardson-D'Jakonov inner iteration.
We use numerical simulation 3 to compare the performance of the direct and iterative solvers. 
\Cref{convergence and comparison} shows that, given a specific $tol$ value (see \cref{adaptive time-stepping}), the accuracy of the direct and iterative solvers are identical.
\cref{fig: direct-iterative comparison} shows, clearly, that the iterative solver is twice as fast as the direct solver.

\cref{fig: direct-iterative comparison} also gives an in-depth look at the inner iterations of the iterative solver.
It shows a positive correlation between $tol$ values and number of inner iterations required.
This was expected as lower $tol$ values force the time-stepper to take smaller steps, which means the solution to the Jacobian update equation \cref{eq: Jacobian update} is not far from the initial guess ${\Delta \boldsymbol{x}}_{0}^{i+1} = \boldsymbol{0}$.
This allows the iterative solver to converge faster for smaller $tol$ values .

\section{Conclusion}\label{conclusion}

In this work, we presented a detailed model and simulation of the retina, which takes into account the retinal physiology as well as the geometry of the eye.
The model is based on the bi-domain equations and is the first bi-domain retina model which takes into account the full geometry of the eye.
It is a versatile model in the sense that it can be used with any model of transmembrane currents.
This model can be viewed as complimentary to the current retinal models, such as the one proposed by Dokos et al.~\cite{Dokos}, which were mainly concerned with electrode stimulation of and signal propagation through the retina.
We detailed how we discretize the model's system of PDEs and our implicit time-stepping scheme which used BDF2 and a Newton-iterative method.
We demonstrated how our adaptive time-stepper was used to significantly decrease simulation time while maintaining accuracy.
We also generalized our bi-domain model to a multi-domain model, which can account for all types of known photoreceptors.

A limitation of this model is that it can only account for one type of neurons present in the retina, namely the photoreceptors.
Adding other types of neurons will increase the scope of applicability of this model since it will reproduce more features of the retina (for example, the b-wave mentioned in \cref{general observations}).
Another limitation of the model is the computational cost of this simulation.
This, unfortunately, has been the downfall of cardiac tissue models using the bi-domain equations as well, which limited their usage~\cite{Keener}.
We hope to address both limitations in a future publication.

An exciting area of application for this model is in aiding electroretinogram (ERG) diagnostics.
The model can be manipulated to mimic many diseases that ERGs are used to diagnose.
Thus, through various techniques such as parameter fitting, we hope to be able to use this model to replicate ERG measurements and subsequently aid ERG diagnostics.

\appendix
\section{Simulation Details}\label{appendix}
\begin{figure}
\includegraphics{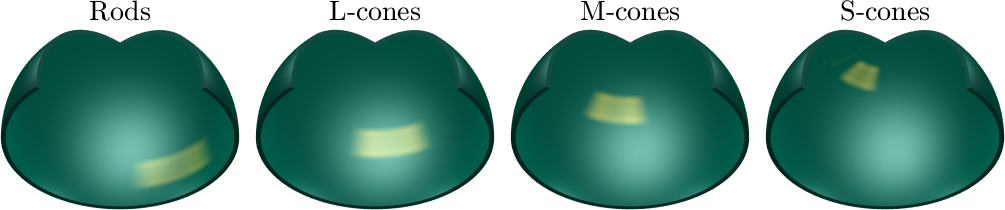}
\caption{Left to right: Location of light stimuli flashed at rods, L-cones, M-cones, and S-cones, respectively. The times of the $20$ ms light pulses were at $t = 0,0.5,1.0,1.5$ s for rods, L-cones, M-cones, and S-cones, respectively}
\label{light stimuli}
\end{figure}

\phantom{hi} \\
{\bf Numerical Simulation 1:}
\begin{itemize}
\item Grid size ($r\cross \theta \cross \varphi$): $30 \times 29 \times 27$.
\item Total number of unknowns: $142352$.
\item Simulation interval: $[0,5]$ (seconds).
\item Active Domains: L-cones.
\item Stimulus: Spatially Gaussian ($\sigma = 50$), $20$ ms light pulses at $t=0, 2.0, 2.1,$ \\ $2.2,\ldots,2.9$ s aimed at center of retina.
\end{itemize}
\noindent
{\bf Numerical Simulation 2:}
\begin{itemize}
\item Grid size ($r\cross \theta \cross \varphi$): $30 \times 35 \times 46$.
\item Total number of unknowns: $1146412$.
\item Simulation interval: $[0,5]$ (seconds).
\item Active Domains: Rods, L-cones, M-cones, and S-cones.
\item Stimulus: Spatially-disjoint, $20$ ms light pulses were given at $t = 0,0.5,1.0,1.5$ s to rods, L-cones, M-cones, and S-cones, respectively (see \cref{light stimuli}).
\end{itemize}

\noindent
{\bf Numerical Simulation 3:}
\begin{itemize}
\item Grid size ($r\cross \theta \cross \varphi$): $30 \times 29 \times 27$.
\item Total number of unknowns: $142352$.
\item Simulation interval: $[0,1]$ (seconds).
\item Active Domains: L-cones.
\item Stimulus: Spatially Gaussian ($\sigma = 50$), $1$ s light flash at $t=0$ s aimed at center of retina.
\end{itemize}
\noindent
{\bf Numerical Simulation 4:}
\begin{itemize}
\item Grid size ($r\cross \theta \cross \varphi$): $30 \times 29 \times 27$.
\item Total number of unknowns: $142352$.
\item Simulation interval: $[0,5]$ (seconds).
\item Active Domains: L-cones.
\item Stimulus: Spatially Gaussian ($\sigma = 50$), $20$ ms light pulse at $t=0$ s aimed at center of retina.
\end{itemize}

\section*{Acknowledgments}
We would like to thank Professor Mary Pugh for all her valuable insight in developing the adaptive time-stepper.

\bibliographystyle{siamplain}
\bibliography{references}

\end{document}